\documentclass[fleqn,usenatbib]{mnras}

\usepackage{newtxtext,newtxmath}

\usepackage[T1]{fontenc}
\usepackage{ae,aecompl}
\usepackage{graphicx}	
\usepackage{amsmath}	
\usepackage{amssymb}	

\usepackage{color,ulem}  
\definecolor{deepblue}{rgb}{0,0,0.5}  
\definecolor{deepred}{rgb}{0.6,0,0}   
\definecolor{deepgreen}{rgb}{0,0.5,0} 
\definecolor{darkgreen}{rgb}{0,0.6,0} 

\newcommand{\out}[1]{}  
\newcommand{\cmc}{cm$^{-3}$}
\newcommand{\cmcs}{cm$^{3}$~s$^{-1}$}

\usepackage{graphicx}	
\usepackage{amsmath}	
\usepackage{amssymb}	
\usepackage{tabularx}   
\usepackage{subfigure}

\def\ha{$H\alpha$}

\def\hb{$H\beta$}

\def\hg{$H\gamma$}
\def\oiii{{[O\sc{iii}]}\/}
\def\oiiia{{[O\sc{iii}]}$\lambda$4959\/}
\def\oiiib{{[O\sc{iii}]}$\lambda$5007\/}

\def\niia{{[N\sc{ii}]}$\lambda$6548\/}
\def\niib{{[N\sc{ii}]}$\lambda$6583\/}

\def\siia{{[S\sc{ii}]}$\lambda$6716\/}
\def\siib{{[S\sc{ii}]}$\lambda$6731\/}

\def\oia{{[O\sc{i}]}$\lambda$6300\/}
\def\oib{{[O\sc{i}]}$\lambda$6364\/}

\def\l{$\lambda$}

\def\l5100{$L_{\it 5100}$}

\title[Multi-wavelength observations of the triple-peaked AGN Mrk\,622]{Multi-wavelength observations of the triple-peaked AGN Mrk\,622}

\author[E.~Ben\'{\i}tez et al.]
{E.~Ben\'{\i}tez,$^{1}$\thanks{Email: erika@astro.unam.mx}
I.~Cruz-Gonz\'alez,$^{1}$
J.~M.~Rodr\'iguez-Espinosa,$^{2,3}$
\newauthor
O.~Gonz\'alez-Mart\'in,$^{4}$
C.~A.~Negrete,$^{5}$
L.~Guti\'errez,$^{6}$
E.~Jim\'enez-Bail\'on,$^{6}$
\newauthor
D.~Ruschel-Dutra,$^{7}$
L.~F. Rodr\'iguez,$^{4}$
L.~Loinard$^{4,1}$ and
L.~Binette$^{1}$
\\
$^{1}$Instituto de Astronom\'{\i}a, Universidad Nacional Aut\'onoma de M\'exico, Ciudad Universitaria, A.~P. 70-264, CDMX 04510, Mexico\\
$^{2}$Instituto de Astrof\'isica de Canarias (IAC), V\'ia L\'actea, s/n, 38205, La Laguna, Spain\\
$^{3}$Departamento de Astrof\'isica, Universidad de La Laguna (ULL), 38205, Spain\\
$^{4}$Instituto de Radioastronom\'ia y Astrof\'isica , Universidad Nacional Aut\'onoma de M\'exico, A.~P. 3-72 (Xangari), 8701, Morelia, Mexico\\
$^{5}$CONACYT Research Fellow - Instituto de Astronom\'{\i}a, Universidad Nacional Aut\'onoma de M\'exico, A.~P. 70-264, CDMX 04510, Mexico\\
$^{6}$Instituto de Astronom\'ia, Universidad Nacional Aut\'onoma de M\'exico, Apdo. Postal 877, Ensenada, 22800 Baja California, Mexico\\
$^{7}$Departamento de F\'isica, Universidade Federal de Santa Catarina, P.O. Box 476, 88040-900, Florian\'opolis, SC, Brazil\\ 
}

\date{Accepted XXX. Received YYY; in original form ZZZ}

\pubyear{2019}

\begin{document}
\label{firstpage}
\pagerange{\pageref{firstpage}--\pageref{lastpage}}
\maketitle


\begin{abstract}

A detailed multi-wavelength study of the properties of the triple-peaked AGN Mrk\,622 showing different aspects of the nuclear emission region is presented. Radio, near- and mid-infrared, optical and X-ray data has been considered for the analysis. In the optical, the WHAN diagnostic diagrams show that the three nuclear peaks are strong active galactic nuclei since the EW of $H{\alpha}$ is $>$\,6 \AA\, and $\log$ [NII]$\lambda$6584/H$\alpha$\, ratio is $>$\,-0.4. Optical variability of both the continuum flux and  intensity of the narrow emission lines is detected in a time-span of 13 years. The size of the narrow line region is found to be 2.7\,pc, with a light-crossing time of 8.7\,y. Analysis done to an archival Hubble Space Telescope image at 1055.2\,nm shows that the host galaxy has a 3.6\,kpc inner bar with PA\,=\,74$^\circ$, faint spiral arms and a pseudobulge, evolving through secular processes. High resolution mid-infrared images obtained with the \textit{Gran Telescopio Canarias (GTC)} and the instrument \textit{CanariCam} show that the nuclear emission at 11.6 $\mu$m is not spatially resolved. Very Large Array archival observations at 10\,GHz reveal a core source with a total flux density of 1.47\,$\pm$\,0.03\,mJy. The spectral index of the core between 8 and 12\,GHz is -0.5\,$\pm$\,0.2, characteristic of AGN. The core deconvolves into a source with dimensions of 82\,$\pm$\,13\,mas\,$\,\times\,$\,41\,$\pm$\,20\,mas, and a  PA\,=\,70\,$\pm$\,18\,deg; which suggests that the core is elongated or that it is constituted by multiple components distributed along a $\sim$65$^\circ$ axis. 

\end{abstract}





\section{Introduction}

Galaxy mergers have been predicted to be the sites of major accretion episodes. Simulations show that they induce strong inflows of gas towards the central region of galaxies via loss of angular momentum.  They are also predicted to play a major role in the growth of super-massive black holes (SMBHs) with masses in the range of 10$^{6}$-10$^{10}$\,M$_\odot$ and in the triggering of Active Galactic Nuclei (AGN) \citep[e.g.,][]{2015ApJ...814..104K,2014MNRAS.441.1297S,2011MNRAS.418.2043E}. 
So-called major mergers are found to be associated with very luminous AGN, i.e., those with L\,$\geq$\,10$^{46}$\,erg\,s$^{-1}$, and are thought to be responsible of the growth of SMBHs that have M$_{BH}\,\geq\,10^{8}\,M_\odot$. Major mergers can also trigger central bursts of star formation (SF) and produce a remnant spheroidal galaxy or a galaxy with a large bulge \citep[][]{1991ApJ...370L..65B,2007Sci...316.1877M,2008ApJ...679..156H}. 
Moreover, semi-analytic models predict that low to intermediate luminosity AGN, i.e., Sy galaxies with L\,$\sim$\,10$^{42}$-10$^{44}$\,erg\,s$^{-1}$, evolve via minor mergers. In this case, the interaction consists of a massive galaxy capturing a gas-rich dwarf galaxy (a satellite galaxy) that is capable of triggering and fueling the AGN activity \citep[see][]{1995ApJ...448...41H,2001A&A...367..428A,2014MNRAS.437.3373N}.

Since it is well known that SMBHs are commonly found in the centre of low redshift  massive galaxies \citep[see][]{1995ARA&A..33..581K,2005MNRAS.364..407C,2005SSRv..116..523F}, it is expected that a merger system should contain two approaching SMBHs. The so-called ``dual active galactic nuclei'' (hereafter, dual-AGN) are proposed to be the result of galaxy merging, with each one having an actively accreating SMBH in its center, with a typical size separation of the order of $\sim$\,kpc \citep[e.g.,][]{2007ApJ...660L..23G,2009ApJ...705L..20X}.  Dual-AGN correspond to a phase in the evolution of the merger, which is predicted to last $\sim$100\,Myr before coalescence takes place \citep{1980Natur.287..307B}. Recent simulations show that minor and major mergers results in dual-AGN lasting in such phase during $\sim$\,20\,-\,70\,Myr and $\sim$\,100\,-\,160\,Myr, respectively \citep{2017MNRAS.469.4437C}. During the dual-AGN phase, both the AGN and the starburst (SB) activity are found to be more vigorous \citep[e.g.][]{2012ApJ...748L...7V,2013MNRAS.429.2594B}.

During the AGN accretion process, the majority of the SMBHs are found to be obscured  by large amounts of gas and dust. There is evidence indicating that the absorbing material surrounding the SMBHs increases during the later phases of the merging sequence \citep[e.g.,][]{2017MNRAS.468.1273R,2017ApJ...848..126S,2015A&A...573A.137L}.  It has  also been found that the fraction of Compton-Thick AGN  (CT; where N$_{H}\,\ge\,10^{24}$\,cm$^{-2}$) is larger in late mergers. Recent studies  of CT AGN support the idea that obscured AGN are connected with merger activity  \citep[e.g.,][]{2015ApJ...814..104K,2016ApJ...825...85K,2017MNRAS.468.1273R,2018Natur.563..214K}.

In recent years, very few kpc-scale separation dual-AGN have been confirmed through X-ray, optical or radio observations 
\citep[e.g.,][]{2003ApJ...582L..15K,2011ApJ...735L..42K,2015ApJ...806..219C,2018MNRAS.480.1639D,2015ApJ...813..103M,2019MNRAS.484.4933R}. 
The separation of the two SMBHs in dual-AGN defines the evolutionary state of the system.  If the two SMBHs are separated by $\sim$\,pc scales they are denoted as binary AGN since they are now forming a bound pair. Some authors suggest that dual and binary AGN are systems with separations $\leq$\,10\,kpc and $\leq$100\,pc, respectively \citep[e.g.][]{2014arXiv1402.0548B}.

To confirm that an object is a dual or binary AGN, high-resolution imaging and spectroscopy observations are required. Up to recently, only one object is considered to be a confirmed binary AGN that has two SMBHs separated by 7\,pc\, \citep[][]{2006ApJ...646...49R}. A few candidates of binary SMBHs have since been suggested by  \citep[e.g.,][]{2017NatAs...1..727K,2017MNRAS.465.4772R}. Insofar as dual-AGN, very few have been confirmed \citep[see][]{2019MNRAS.484.4933R}. Identifying that an AGN is binary is crucial for understanding the final stages of galaxy mergers. In particular, binaries separated by less than one parsec are expected to emit gravitational waves (GW), leading to the final binary SMBHs coalescence \citep[e.g.,][]{2014SSRv..183..189C}.  

Initially, finding AGN that show double or triple peak narrow emission lines in their spectra was considered as evidence for harboring dual or binary AGN \citep[][]{2009ApJ...705L..76W,2010ApJ...708..427L,2010ApJ...716..866S,2011ApJ...727...71F}.  However, double-peaked emission lines can instead be tracing AGN with peculiar kinematics, such as  biconical outflows, winds, jet-cloud interactions  \citep[e.g.,][]{2000ApJ...532L.101C,2009ApJ...705L..20X,2011ApJ...727...71F,2017MNRAS.465.4772R,2017ApJ...846...12K,2019ApJ...873...11J}. Hence, double-peaked narrow emission lines are rarely found to be due to a dual 
or binary AGN system\citep[e.g.,][]{2016ApJ...832...67N,2018ApJ...867...66C} and follow-up observations are therefore necessary to confirm the existence of a SMBHs pair among the available candidates.

It is however important to mention that the physics related to the AGN triggering mechanism is still under debate \citep[e.g.,][and references therein]{2019arXiv190508830E}. Some studies find that AGN triggering may instead be linked to the host galaxy evolution\,
\citep[e.g.,][]{2005Natur.433..604D,2015MNRAS.447.2123C,2017MNRAS.469.4437C}. In particular, low to intermediate luminosity AGN are proposed to evolve through secular processes, internal to the host galaxies, which last longer than the dynamical timescales \citep[see][]{2019NatAs...3...48S,2014MNRAS.445..823H}. Secular evolution generates bar-like structures, which live longer than the mergers, can also drive gas to the inner regions, creating a pseudobulge which has a central black hole mass in the range of  M$_{BH}$\,=\,(1-3)$\,\times\,$10$^{7}M_{\sun}$ \citep[e.g.,][]{2014MNRAS.445..823H,2009ApJ...694..599H}. Minor mergers, however, that involve satellite accretion, do not damage the galactic disks \citep[see][]{2001A&A...367..428A}. 

Our target AGN, Mrk\,622, was recently found to show evidence, at optical wavelengths,  of triple-peaked narrow emission lines. The analysis of this Seyfert\,2 galaxy (Sy\,2) was done with data obtained with the William Herschel Telescope (\textit{WHT}). Mid-resolution spectra showed that the central peak is due to a SB+AGN source, while the blue and red-shifted components occupy the locus of AGN in the Baldwin, Phillips and Terlevich diagnostic diagrams \citep[BPT;][]{1981PASP...93....5B}. These results along with the spatial separation of the line peaks have been interpreted by \citet{Benitez2018} as evidence for a candidate binary AGN being present in Mrk\,622. 

In the current follow-up work, a multi-wavelength study of Mrk\,622 is presented that is based on several observing facilities. Since the possibility of a binary AGN has to be tested via observations of both nuclei, all the available \textit{Hubble Space Telescope (HST)} archival NIR data have been carefully analysed. Furthermore, mid-IR observations obtained with \textit{CanariCam} attached to the \textit{Gran Telescopio Canarias (GTC)} are presented and included in our analysis. Finally, archival X-ray \textit{(XMM-Newton)}, mid-infrared \textit{(Spitzer/IRS)} spectroscopy and  radio continuum \textit{(VLA)}  observations are also included to complete our study. 

Since this work is a multi-wavelength study, at the end of each section or subsection presented below, a physical explanation of the observations is given.
The discussion section will function to incorporate all of these individual discussions into one. Section~\ref{spectral} deals with the analysis of the optical spectra obtained with both the \textit{WHT} and the \textit{SDSS} telescopes; section ~\ref{phot} covers the analysis of the NIR 
\textit{HST} image, while section ~\ref{mir-obs} presents the mid-infrared data, followed by section ~\ref{xray}, which goes over the X-ray and mid-infrared data analysis. Section ~\ref{VLA} shows the radio maps analysis and section ~\ref{res} is dedicated to an overview of our results while a final discussion is presented in section ~\ref{dis}.

Throughout the paper we adopt a cosmology where 
$H_{0}$ = 69.6\,km\,s$^{-1}$\,Mpc$^{-1}$, $\Omega_{m}$ = 0.286 and $\Omega_{\lambda}$ = 0.714 \citep{2014ApJ...794..135B}.

\section{Optical Spectroscopy data Analysis}
\label{spectral}

\citet{Benitez2018}, hereafter Paper\,I,  carried out a careful analysis of the optical spectra. These authors showed that Mrk\,622 has triple peaked narrow emission lines in data obtained with the Intermediate dispersion Spectrograph and Imaging System (ISIS), attached to the William Herschel Telescope (\textit{WHT}) on January 28 2015. The same results were obtained when analysing the Sloan Digital Sky Survey (\textit{SDSS-DR7}) spectrum of Mrk\,622  on December 11 2001. All the lines in both spectra were resolved into  blue-shifted, red-shifted and systemic velocity (central) components. Note that the time-span between the \textit{SDSS-DR7} and the \textit{WHT} spectra is somewhat larger than 13 years. 

In Paper\,I, the \textit{WHT} spectrum shows that between the blue and red-shifted components, the angular projected spatial offset is $\sim$0.81\,$\pm$\,0.4 pixels or $\sim$0.16\,\arcsec. This offset was calculated using the \textit{IRAF/IMCENTROID} routine. The derived spatial separation at the distance of Mrk\,622 between the blue and red components, both assumed as AGN, is $\sim$76~pc (c.f., Figure~2 in Paper\,I and this work in Figure~\ref{fig:multi}(b)).

Figure~\ref{fig:Fig1} shows the spectra from both telescopes, with the \textit{WHT} shown in black and the \textit{SDSS} in blue. For comparison purposes, Figure~\ref{fig:Fig2} shows an archive image of Mrk\,622 from the \textit{SDSS-DR7} data set in which a central circle delineates the size of the \textit{SDSS} fiber of 3\arcsec. Overlaid to it is the slit used with \textit{ISIS} at the \textit{WHT}, which has a size of 4\arcsec\,$\,\times\,$\,1\arcsec. The position angle of the \textit{ISIS} slit was PA$\sim$\,-\,92$^{\circ}$. Since the aperture used to extract the spectra from the \textit{WHT} observations was chosen to be 3\arcsec, both telescopes present an equivalent spatial coverage. However, the actual spatial sampling might slightly differ as there is no way to determine the exact position of either telescope during the observations.  The data from both telescopes were calibrated using spectrophotometric standard stars. The resulting spectra are shown in Figure~\ref{fig:Fig1}. In order to facilitate comparison, the lower panels show three selected spectral sections. The blue spectra were vertically displaced in order that they overlap at the specific wavelengths given in the caption. The important result is that the spectra show clear differences in continuum flux (upper panel) and in the line fluxes of the narrow line region (NLR). These differences, which we presume are not the result of detector positioning error, will be analysed in detail in the next subsections.

\subsection{Emission line fluxes and classification}
\label{class}

In Table~\ref{table:Table1} and Table~\ref{table:Table2}, the line fluxes, the FWHM and the equivalent widths (EW) for the important emission lines are presented, for the \textit{WHT} and \textit{SDSS} spectra, respectively. These parameters were estimated under the assumption that each set of lines is properly fitted using a three Gaussian components model (3G model). The continuum flux and line intensities of Mrk\,622 are found to be more luminous in 2015 with respect to the previous \textit{SDSS} data taken $\sim$13\,yr earlier. Note that all the lines from the \textit{WHT} spectrum have increased in flux with respect to the earlier \textit{SDSS} observations. The dominant source of this flux increment is the central component, although both the red and blue component fluxes have also increased. To complement both  tables, we present in Table~\ref{table:Table3} the H$\gamma$ and [OIII]$\lambda$4363 results on the profile modelling with the 3G model. It is interesting to note that the [OIII]$\lambda$4363 is detected only in the \textit{WHT}. The detection of this weak emission line is crucial, since it enables to estimate the density of the [OIII] emitting gas using the approach of \citet{2005MNRAS.358.1043B}. More details on the method used  are presented in subsection ~\ref{varia} below.

It is also found that the centroid of the red component of the [OI]$\lambda$6300 emission line is displaced by 8.6$\pm$ 0.9\,\AA, i.e., by $\sim$420\,km\,s$^{-1}$ from the systemic velocity (represented by the vertical dashed line in the upper panels of  Figure~\ref{fig:Fig3}) in the \textit{WHT} spectrum. Such is not the case in the \textit{SDSS} spectrum. Moreover, the observed flux ratios between the blue and red components have grown from just 1 (\textit{SDSS}) to a factor 2.2 (\textit{WHT}).

Comparison of the doublet [SII]$\lambda\lambda$6716,6731 line intensities between \textit{WHT} and \textit{SDSS} clearly show significant variations of both the central and red components, see lower panels of Figure~\ref{fig:Fig3}. The observed decrease in the  $\frac{[SII]\lambda{6716}}{[SII]\lambda{6731}}$ ratio, if real, would imply a density increase over time of the emission gas, provided the changes are not the result of on-source off centered observations or of using a slightly different spatial resolution.   

When the line ratios from the \textit{WHT} of the three components are plotted in the BPT diagrams (c.f., Figure~3 in Paper\,I), one finds that both the blue and red-shifted components lie on the AGN locus in each of the three diagrams. On the other hand, the systemic velocity component (i.e., the central component) lies on the Starburst (SB) region in two of the BPT diagrams, while for the [OIII]/H$\beta$ vs. [NII]/H$\alpha$ diagram, it lies within the AGN+SB region. Therefore, the BTP diagrams give us an ambiguous classification for the central component. In BPT diagrams where the line ratios  fall in the locus of AGN+SB, suggests the presence of an old, ionising stellar population  \citep{2008MNRAS.391L..29S} provided that the emission of the $H\alpha$ equivalent width is of order of a few \AA\ \citep{1994A&A...292...13B}.  

The line ratios obtained from both telescopes were shown in Table\,3 of Paper\,I. The corresponding BPT diagnostic diagrams of the intensity ratios estimated from both \textit{WHT} and \textit{SDSS} are now presented in Figure~\ref{fig:Fig4}, which show that the position of the blue and red components are now in close agreement, although a small displacement can be observed between the red components due to the changes in the different line intensities. 

Therefore, in order to verify the ionisation origin of the three components, the WHAN diagram ([NII]$\lambda$6584/H$\alpha$ vs. $EW$(H$\alpha$) ought to be used to classify the line emission from Mrk\,622. Such diagram was proposed by \citep[][]{2011MNRAS.413.1687C} and \citep{2015MNRAS.449..559S} and is the result of the analysis of $\sim$700,000 galaxies of the \textit{SDSS-DR7} data release. \citet{2011MNRAS.413.1687C} have shown that it provides a more reliable classification than the one found using BPT diagrams, since the separation of SF and AGN galaxies is well defined. In the WHAN diagram, strong AGN sources (sAGN) have $\log$\,[NII]/H$\alpha$\,>\,-0.4 and $EW$(H$\alpha$)\,>\,6\,\AA; weak AGN (wAGN) have $\log$\,[NII]/H$\alpha$\,>\,-0.4 and 3$\,<\,EW$(H$\alpha$)$\,<\,$6\,\AA\,; retired galaxies (RG) have $EW$(H$\alpha$)$\,<\,$3\,\AA\,; passive galaxies (line-less galaxies) have both $EW$(H$\alpha$) and $EW$[NII]\,<\,0.5\,\AA\,; and pure Star Formation galaxies (SF) have $\log$\,[NII]/H$\alpha$\,<\,-0.4 and $EW$(H$\alpha$)\,>\,3\,\AA. 

In the last column of Tables~\ref{table:Table1} and ~\ref{table:Table2}, the $EW$(H$\alpha$) values obtained for each component are presented. Using these values, the three peaks of Mrk\,622 were positioned in the WHAN diagram (see Figure~\ref{fig:Fig5}), superimposed to the locations of their proposed classification, which was previously reported by \citet[][]{2011MNRAS.413.1687C} and presented in their Figure\,6 for the SDSS-DR7 galaxies. 

From both spectra, the inferred values for the central, blue and red components are $EW$(H$\alpha$)\,$>$\,6\,\AA\, and $\log$\,[NII]/H$\alpha$\,$>$\,-0.4, indicating that the three components classify Mrk\,622 as a sAGN.

It is important to note that the central component has 
$\log$[NII]/H$\alpha$\,$\sim$\,-0.1 in both the \textit{WHT} and \textit{SDSS} spectra.  For this particular component, in the BPT diagrams its classification is found to be ambiguous (see Figure~\ref{fig:Fig4}), but in the WHAN diagram it is clearly classified as sAGN. 

Using the \textit{WHT} spectrum, the total spatial width of the H$\alpha$ emission was estimated, following the subtraction of the seeing contribution in the spatial direction. The size of the H$\alpha$ emitting region turns out to be 3.26\,kpc. This extended emission region might be produced by a starburst-driven outflow 
\citep[][and references therein]{2005ARA&A..43..769V}, since outflows are driven primarily by the kinetic energy from massive stars.  In a similar way, the size of the [OIII]$\lambda$5007 emission region is found to be 0.90\,kpc. The extended emission of [OIII]$\lambda$5007 might instead be produced by AGN-driven outflows or winds.

\subsection{Optical Variability Analysis}
\label{varia}

In the context of the unified scheme for AGN \citep{1993ARA&A..31..473A}, where all AGN are assumed with strong similarities, Sy\,2 galaxies (Type\,II AGN) are proposed to have a hidden broad line region (BLR) due to obscuration along the observer's line of sight.  Nevertheless, the unified scheme has found problems explaining Sy\,2 that either cannot form a BLR or lacks the surrounding obscuring medium, i.e., like those called true Sy\,2 objects. A distinct way to explain obscured and unobscured AGN arises when variations in the X-ray spectra of AGN that are found to occur in timescales of few years. The variations show changes from Compton-thin to reflection dominated or Compton-thick AGN, and were named as ``changing look'' (CL) AGN \citep[][]{2003MNRAS.342..422M,2005A&A...442..185B}. This, however, is similar to what has been observed in nearby AGN since the 80s in the optical bands, particularly in Seyfert galaxies that change their Sy type,  from Sy\,1 to Sy\,2 or  changes to intermediate Sy types like Sy\,2 to Sy\,1.5, etc. \citep[e.g.,][and references therein]{1984MNRAS.211P..33P,2010A&A...509A.106S,2014ApJ...796..134D,2015ApJ...800..144L,2018ApJ...866..123M,2018A&A...619A.168K}.

\begin{figure*}
\begin{center}
\includegraphics[width=1.9\columnwidth,height=8.5cm]{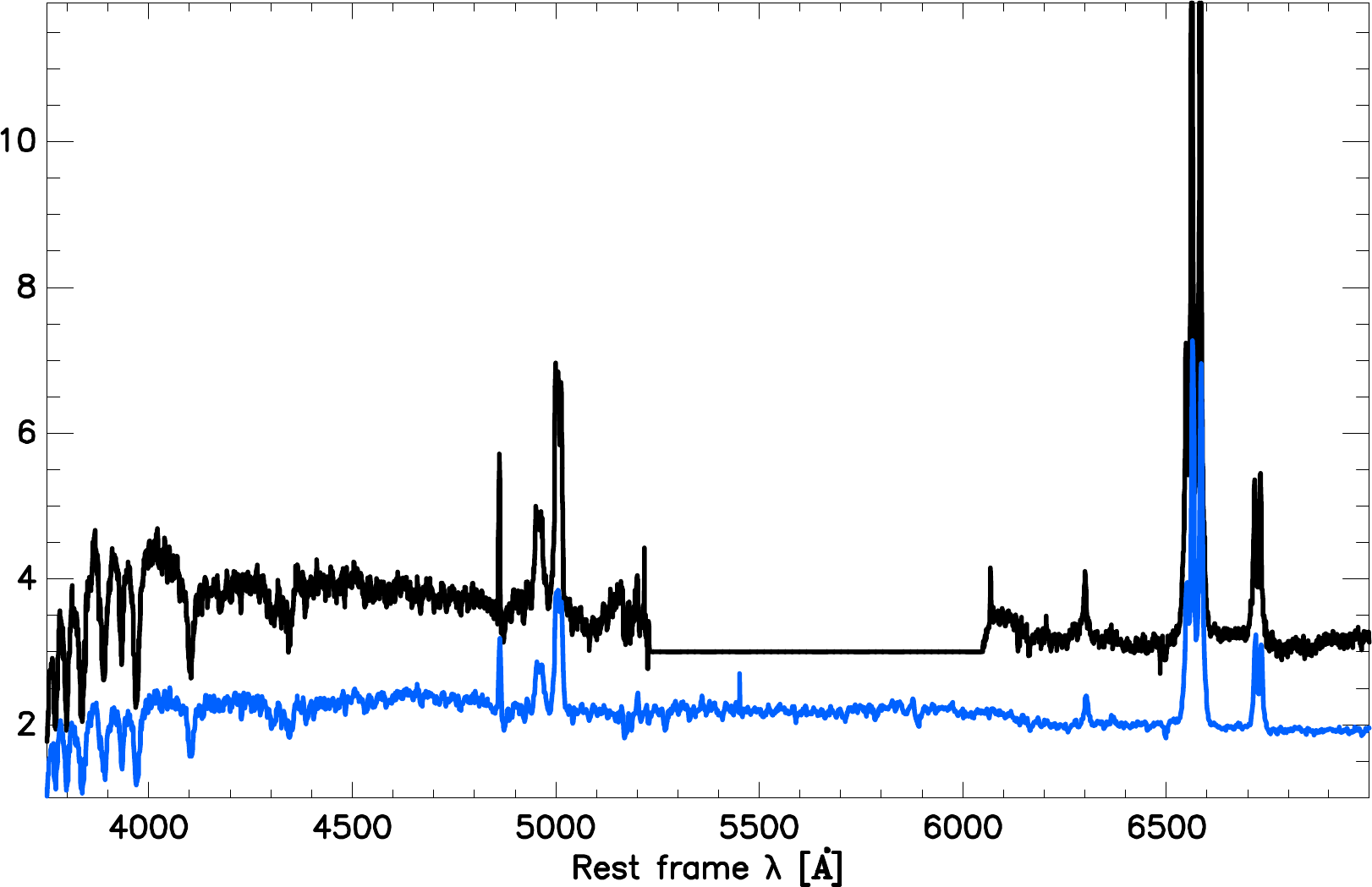}\\
\subfigure{\includegraphics[width=0.65\columnwidth,height=5cm]
{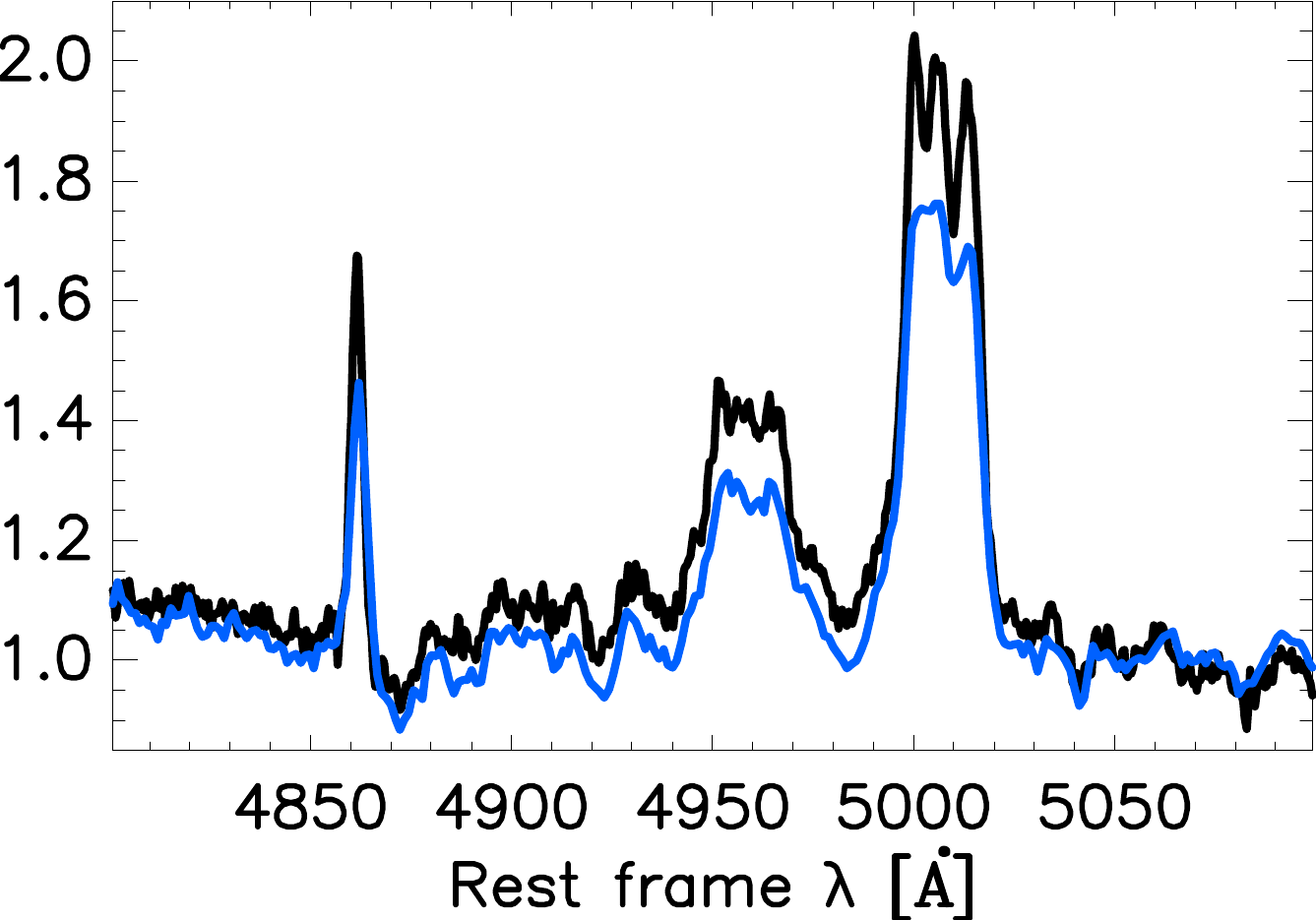}} 
\subfigure{\includegraphics[width=0.65\columnwidth,height=5cm]
{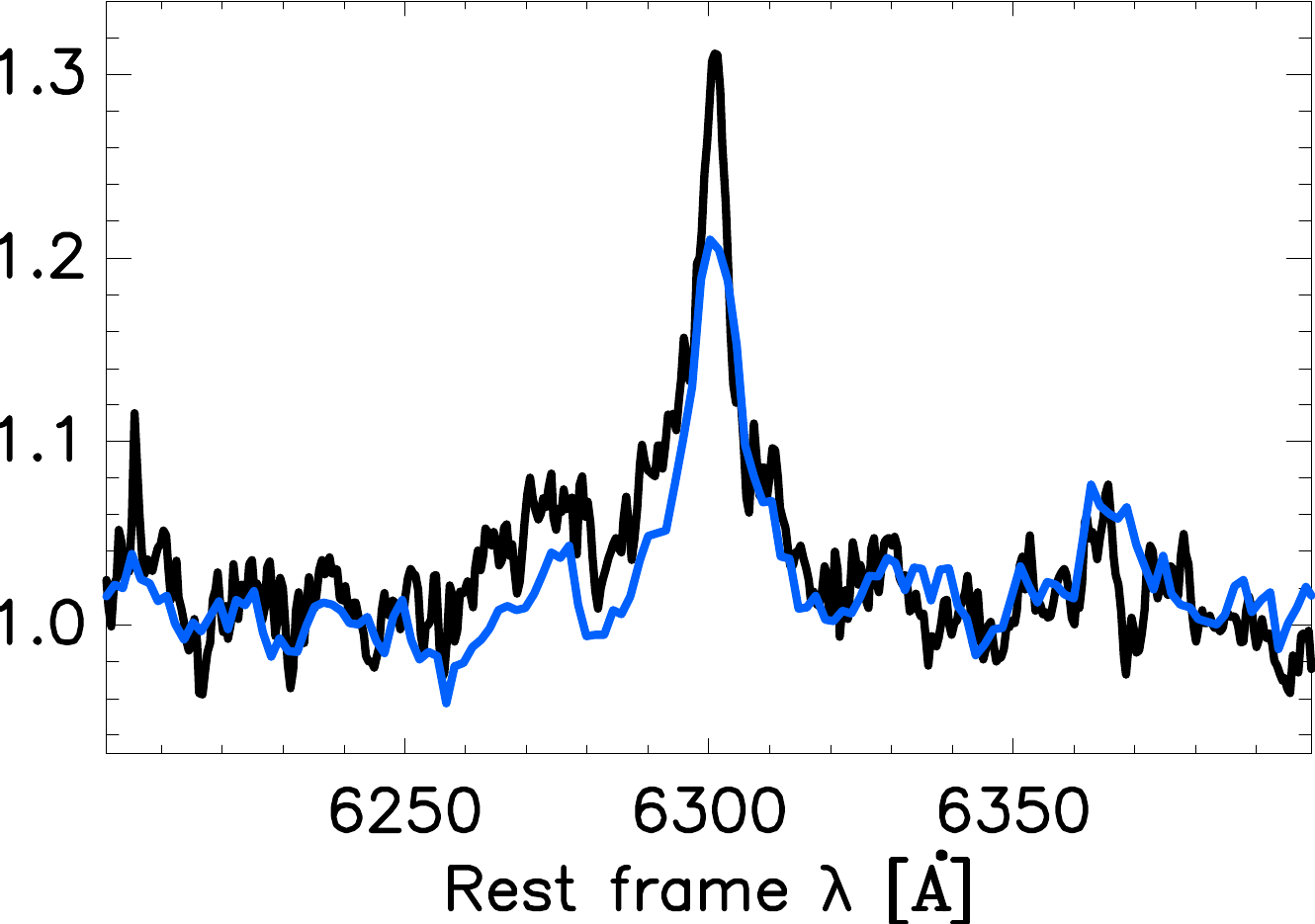}}
\subfigure{\includegraphics[width=0.65\columnwidth,height=5cm]
{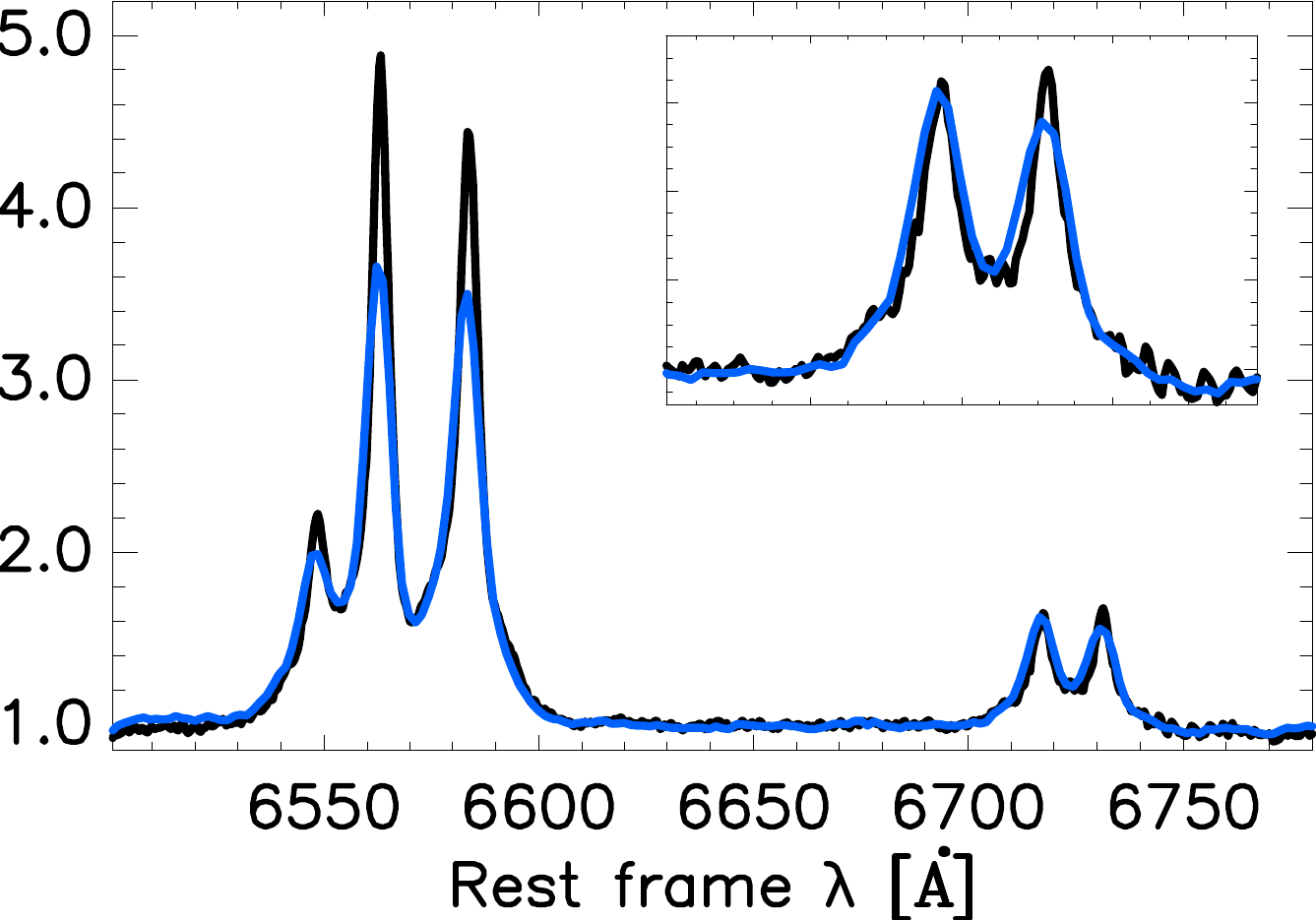}}
\caption{
Upper panel shows the spectra obtained with both the \textit{WHT} and the 
\textit{SDSS} telescopes in units of 10$^{-15}$\,erg\,s$^{-1}$\,cm$^{-2}$\,\AA$^{-1}$. The \textit{WHT} and \textit{SDSS} spectra are shown in black and blue, respectively. For comparison and to show the flux variation in the emission lines, the lower panels display different sections of the spectra that were normalized. The lower left panel shows the [OIII]$\lambda\lambda$4959,5007, and H$\beta$ region, normalized at 5050\AA. The lower central panel shows the [OI]$\lambda\lambda$6300,6364 region normalized at 6350\AA. The lower right panel shows the H$\alpha$ and [SII]$\lambda\lambda$6717,6731 region normalized at 6650\AA. A blow-up of the [SII]$\lambda\lambda$6717,6731 lines shows the variation of the fluxes.}
\label{fig:Fig1}
\end{center}
\end{figure*}

\begin{figure}
\begin{center}
\includegraphics[width=\columnwidth]{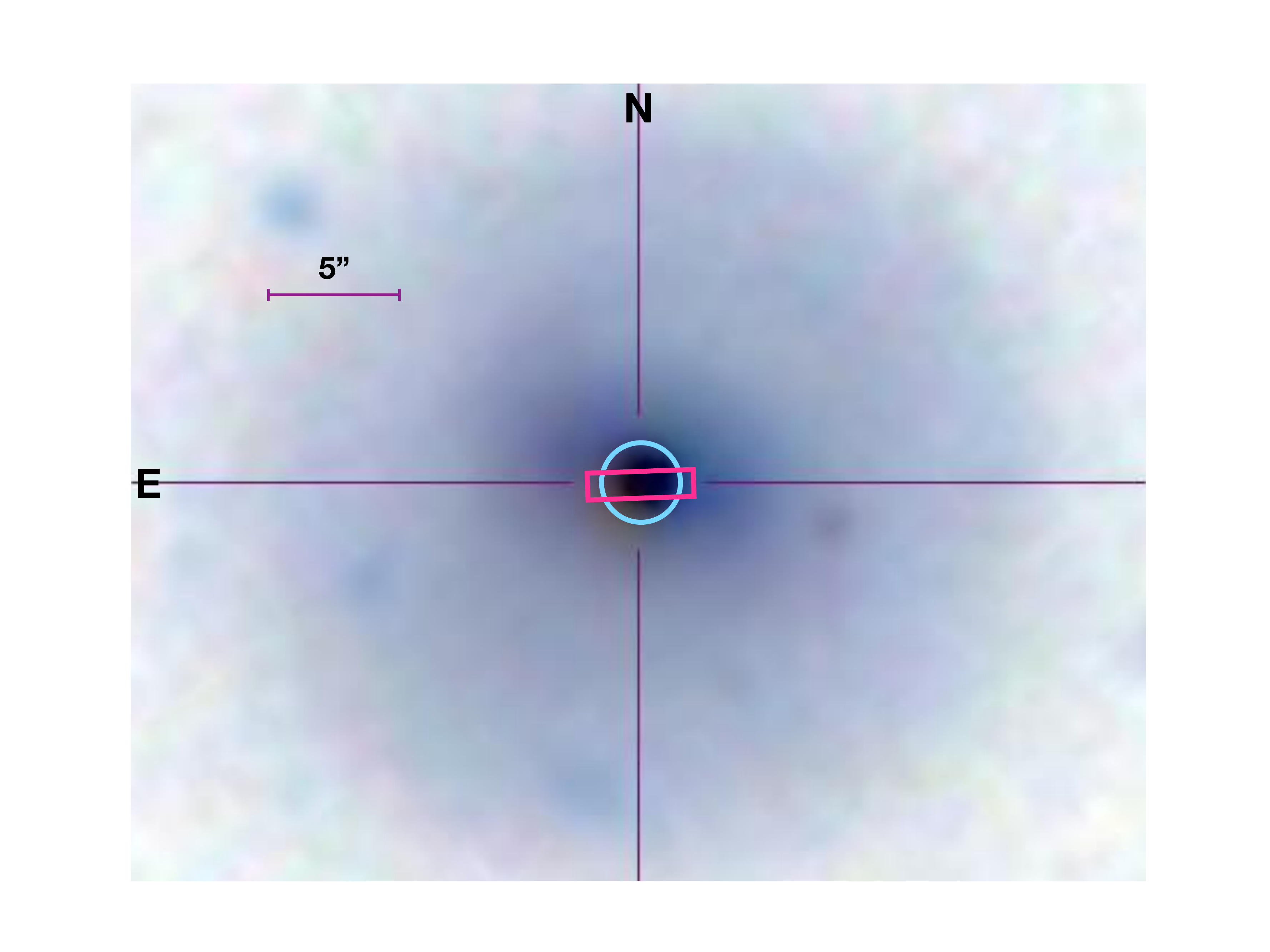}\\
\caption{\textit{SDSS} archival image of Mrk\,622 is shown. The light-blue color circle shows the size of the fiber of the \textit{SDSS} spectrograph (3\arcsec). Overlaid to it, a rectangle in magenta shows the size of the slit used during the observations with the \textit{WHT} (4\arcsec\,$\,\times\,$\,1\arcsec).} 
\label{fig:Fig2}
\end{center}
\end{figure}

\begin{figure*}
\includegraphics[scale=0.4]{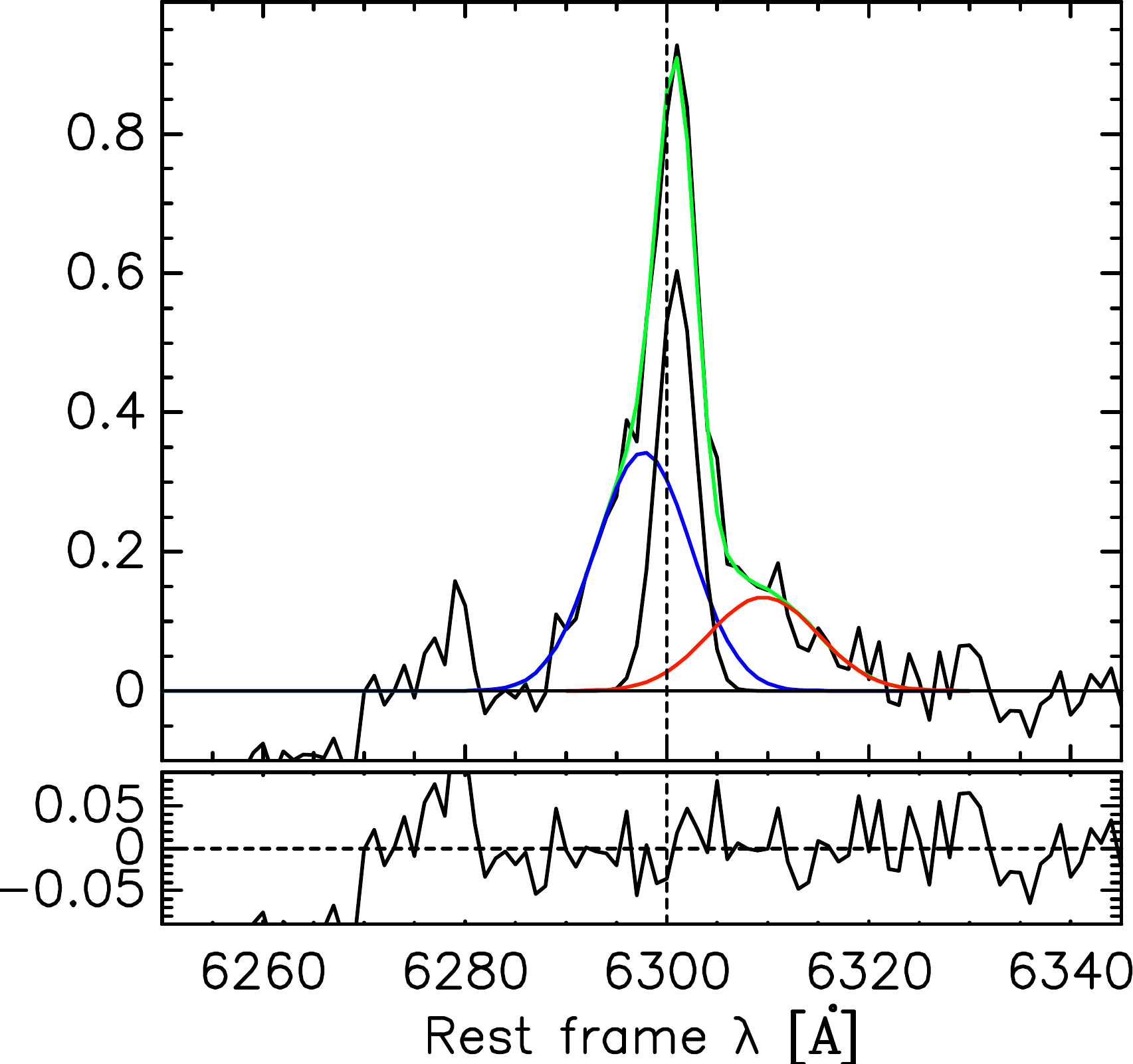}\includegraphics[scale=0.4]{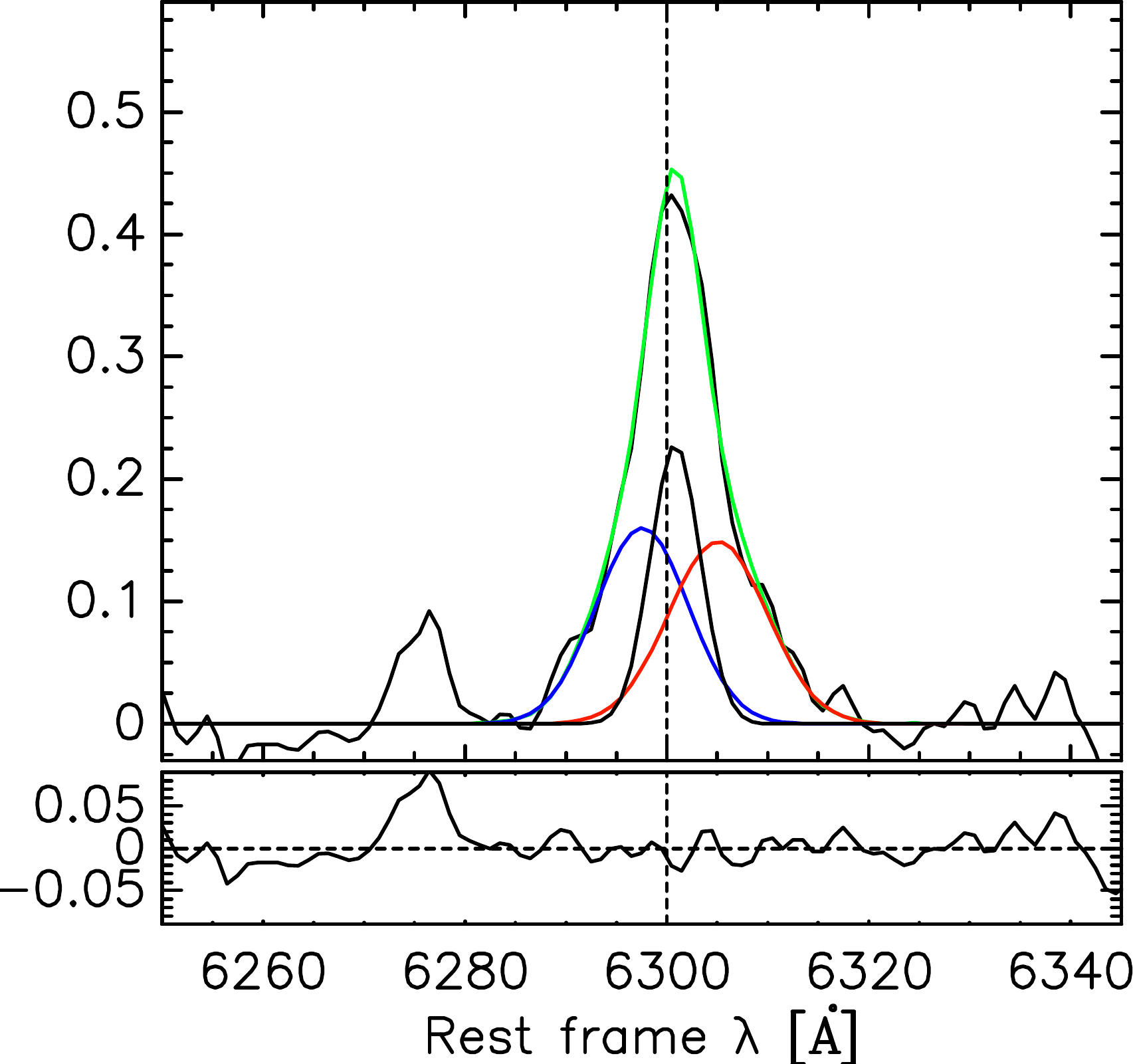}\\
\includegraphics[scale=0.4]{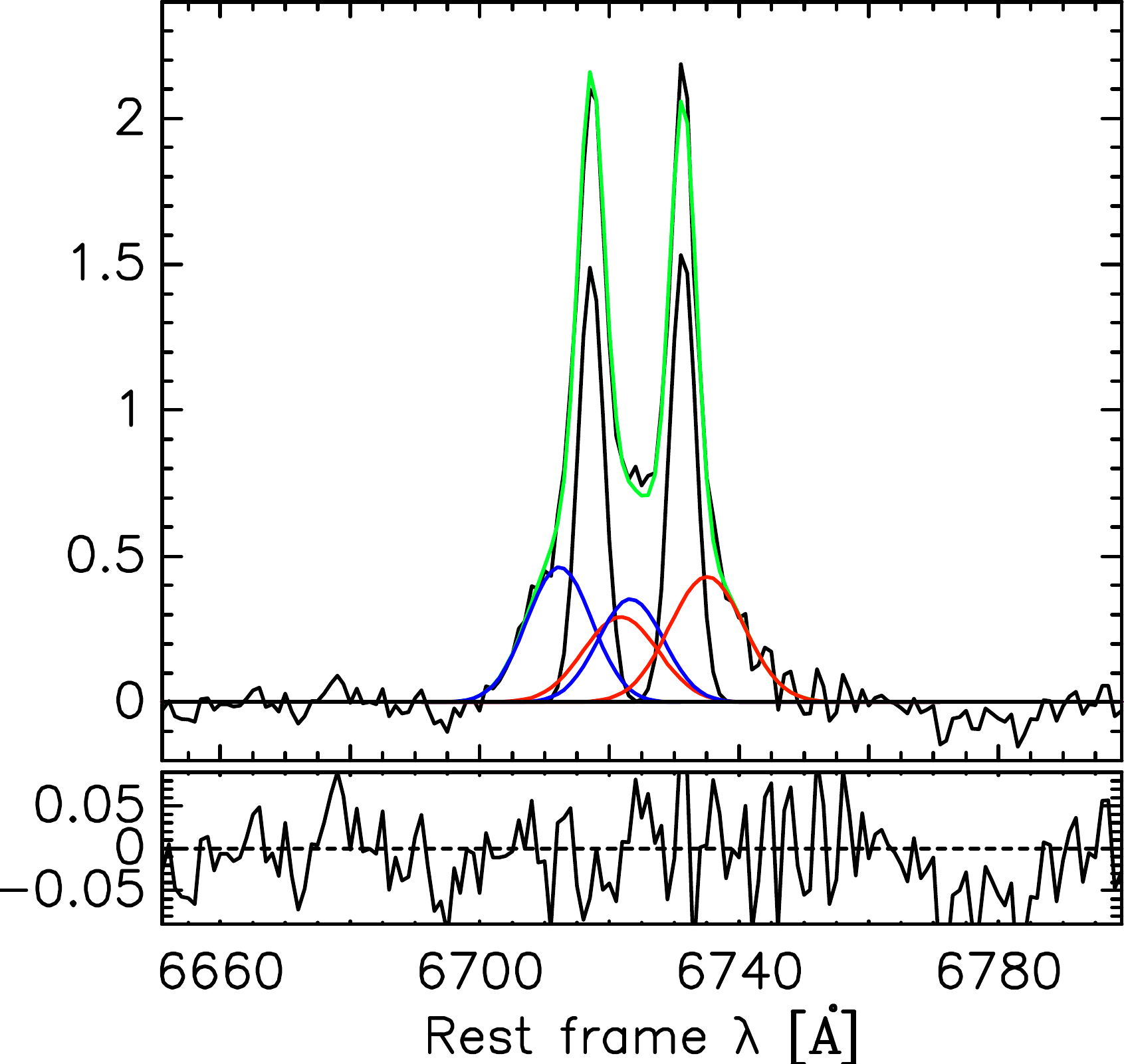}\includegraphics[scale=0.4]{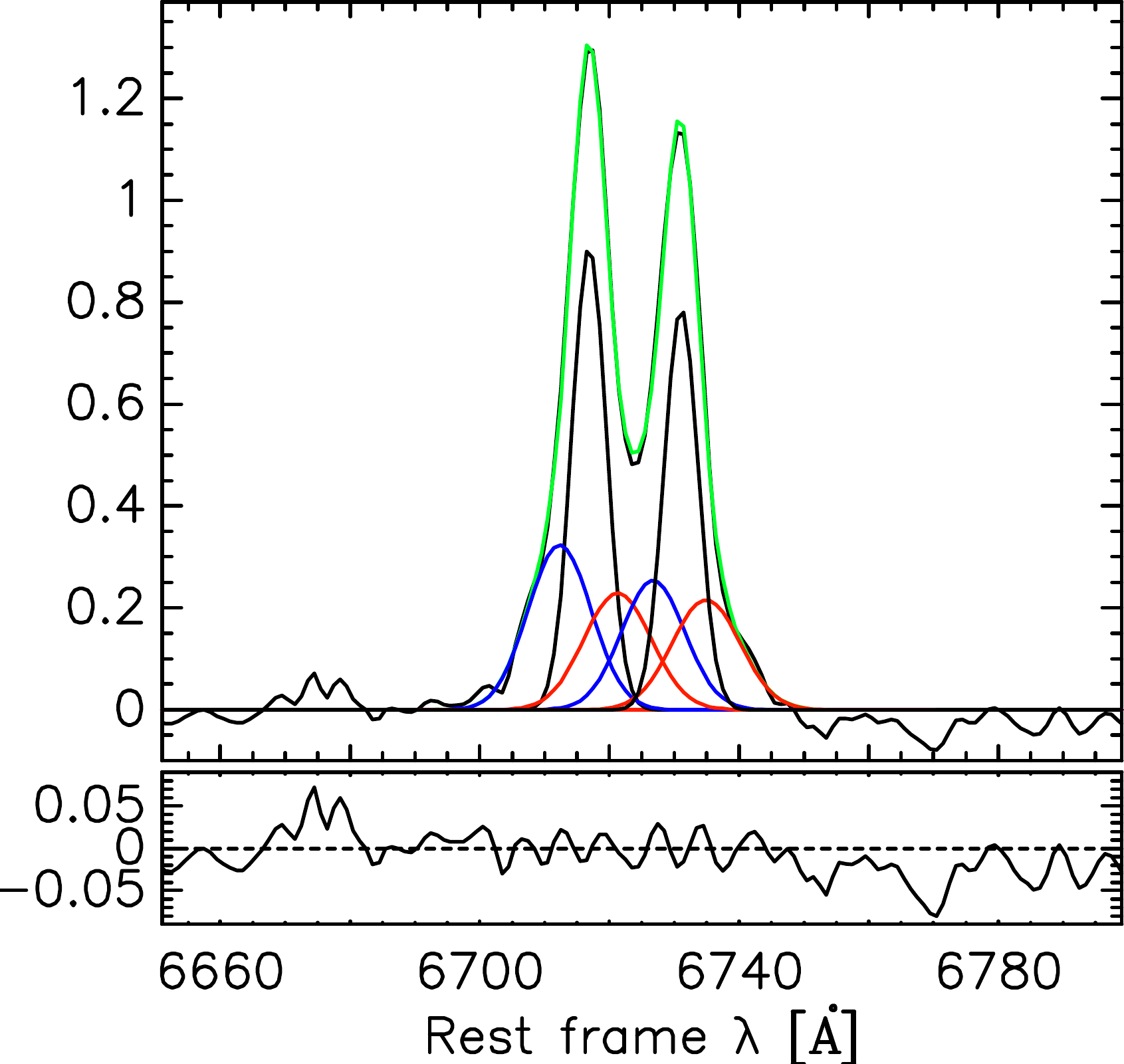}\\
\caption{
Upper panels show the profile fitting done to [OI]$\lambda\lambda$6300,6364 spectral region (left \textit{WHT} spectra; right \textit{SDSS} spectra). 
The dashed line represents the systemic velocity, that is, the centroid of the central component. The lower panels show the profile fitting done to the [SII]$\lambda\lambda$6716,6731 spectral region (left \textit{WHT}; right \textit{SDSS}). In all panels the central component is shown in black, in blue and red the corresponding blue and red-shifted components. Green colour shows the best fit. 
}
\label{fig:Fig3}
\end{figure*}

\begin{figure}
\subfigure{\includegraphics[width=7cm,height=5.5cm]{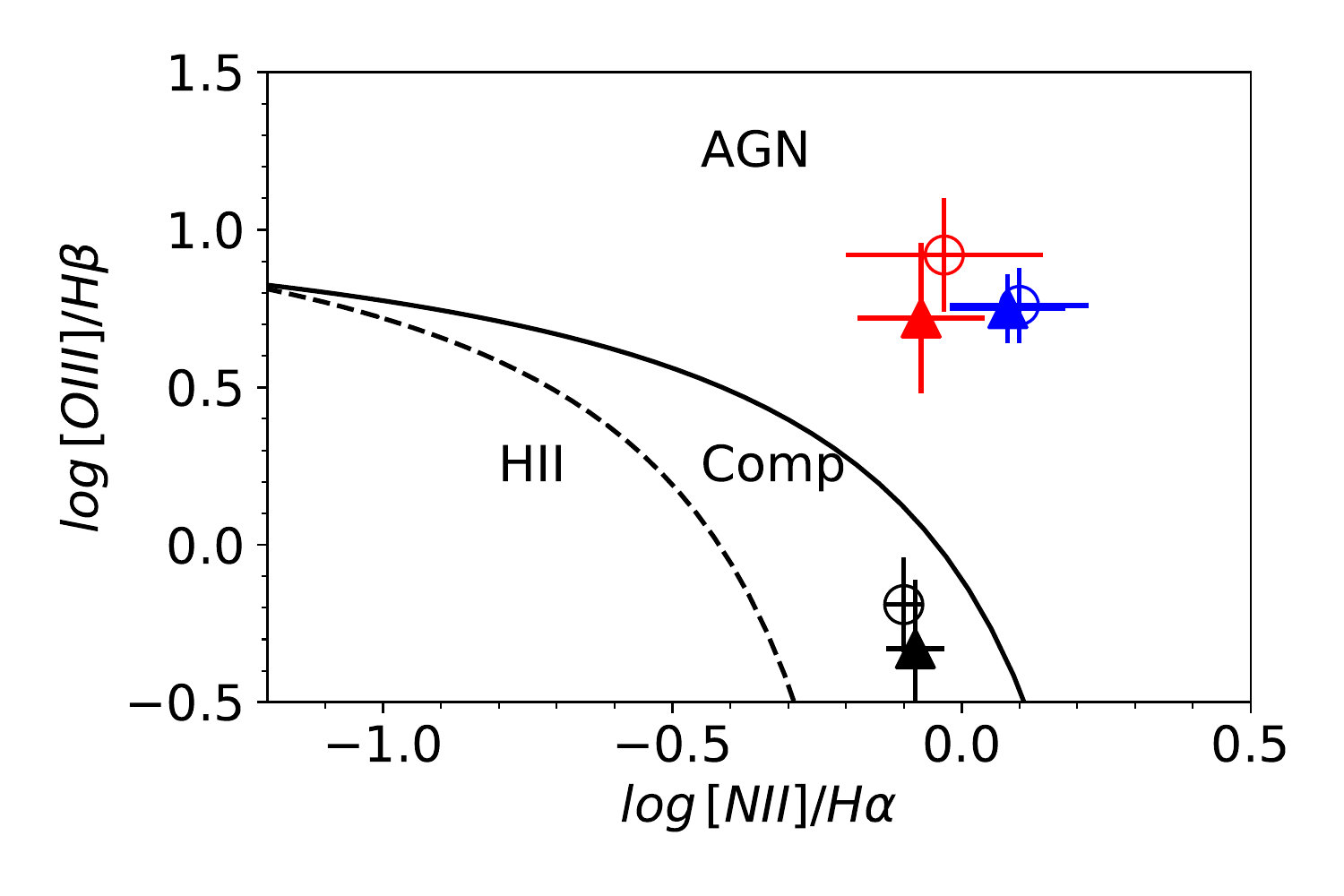}}\\
\subfigure{\includegraphics[width=7cm,height=5.5cm]{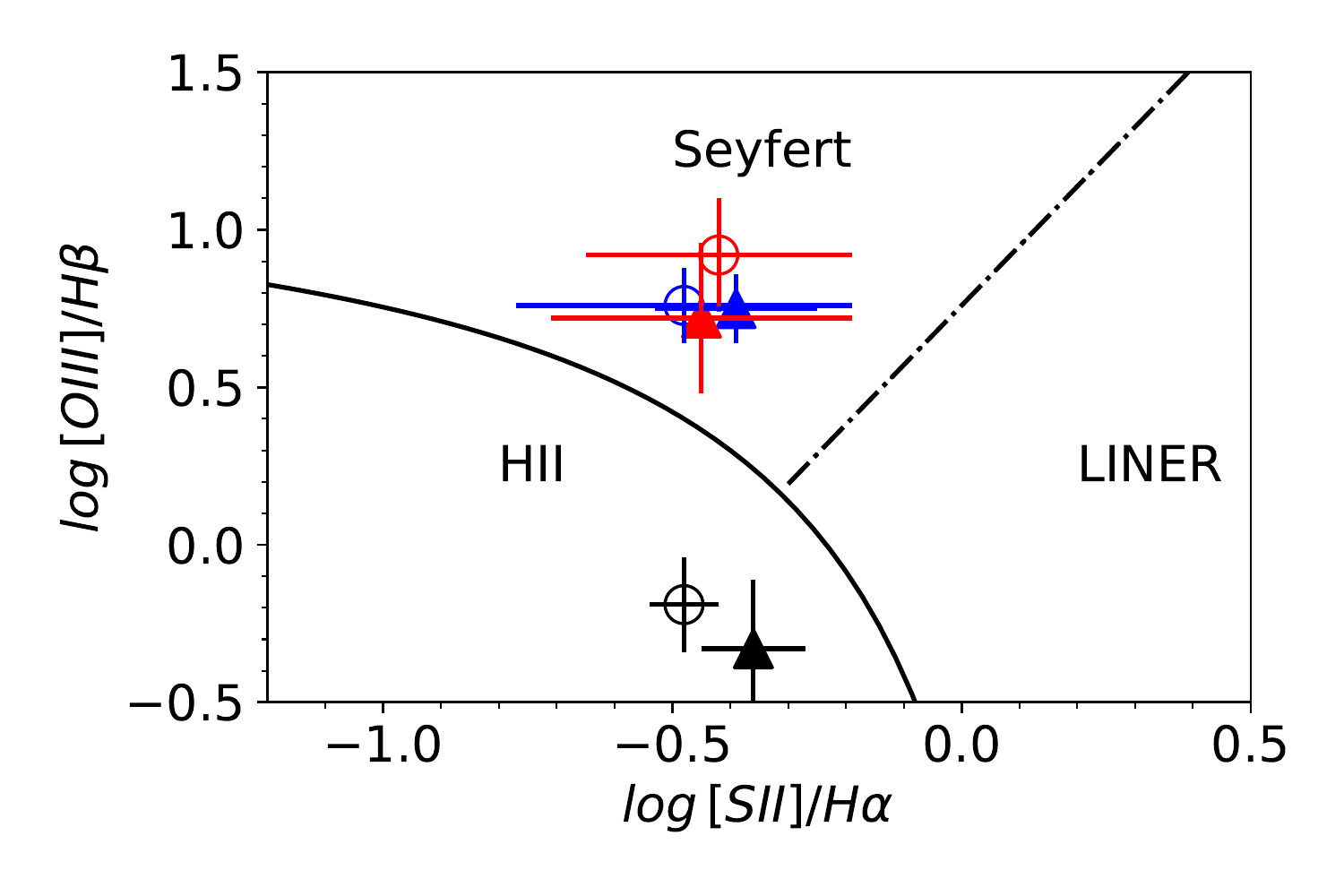}}\\
\subfigure{\includegraphics[width=7cm,height=5.5cm]{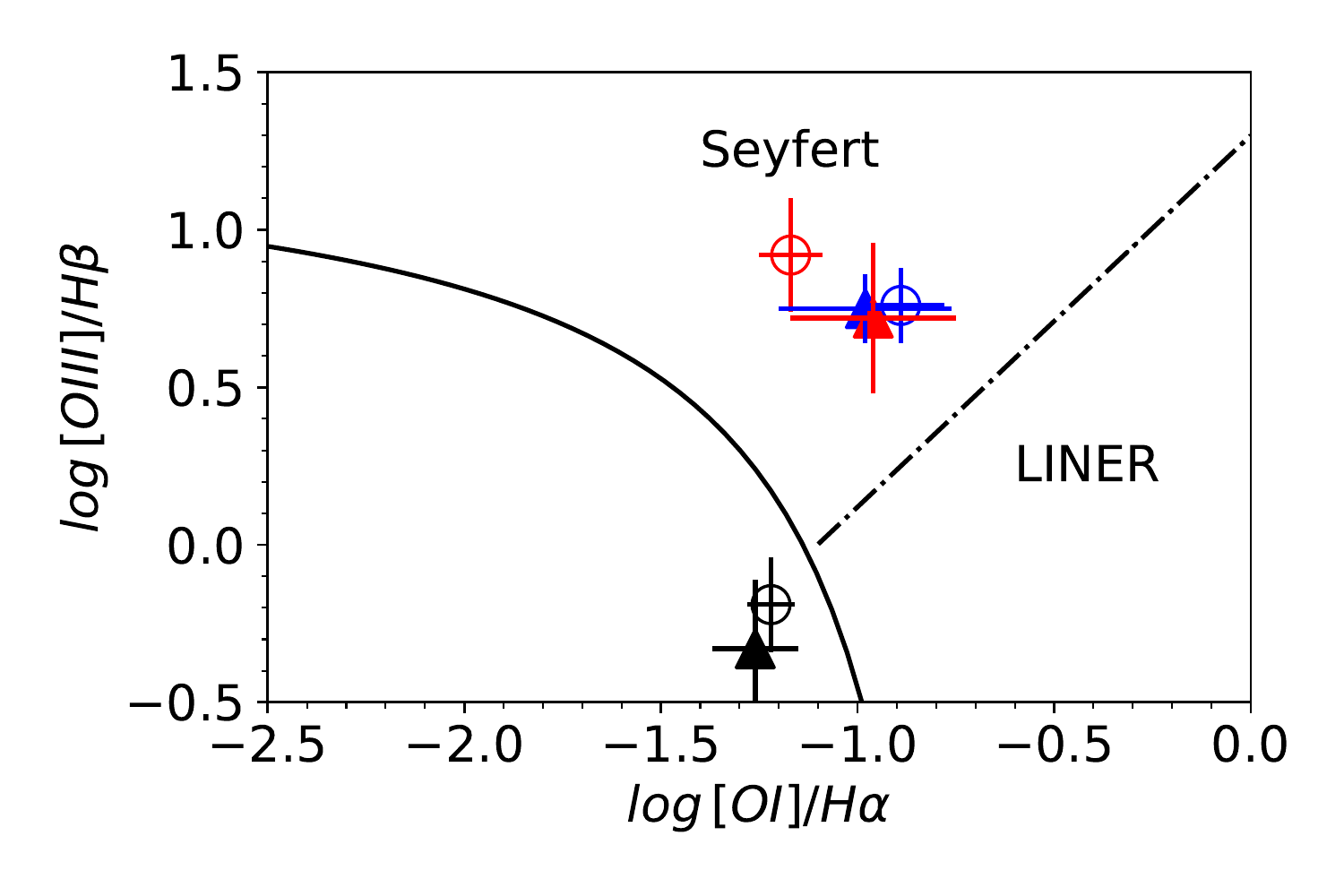}}\\
\caption{\textit{BPT} diagrams. The central, blue and red-shifted components are shown for the \textit{WHT} (open circles) and \textit{SDSS} (triangles) spectra. The solid black curve marks the extreme Starburst classification proposed by \citet{2001ApJ...556..121K}, the dashed line marks the pure star formation region found by \citet{2003MNRAS.346.1055K}, and the dotted-dashed lines divide Seyferts/LINERS accordingly to \citet{2006MNRAS.372..961K}. Blue-shifted, red-shifted and central components are shown in blue, red and black colours.}
\label{fig:Fig4}
\end{figure}

\begin{figure}
\begin{center}

\includegraphics[width=\columnwidth,trim=3.0cm 1.0cm 3.0cm 1.0cm]{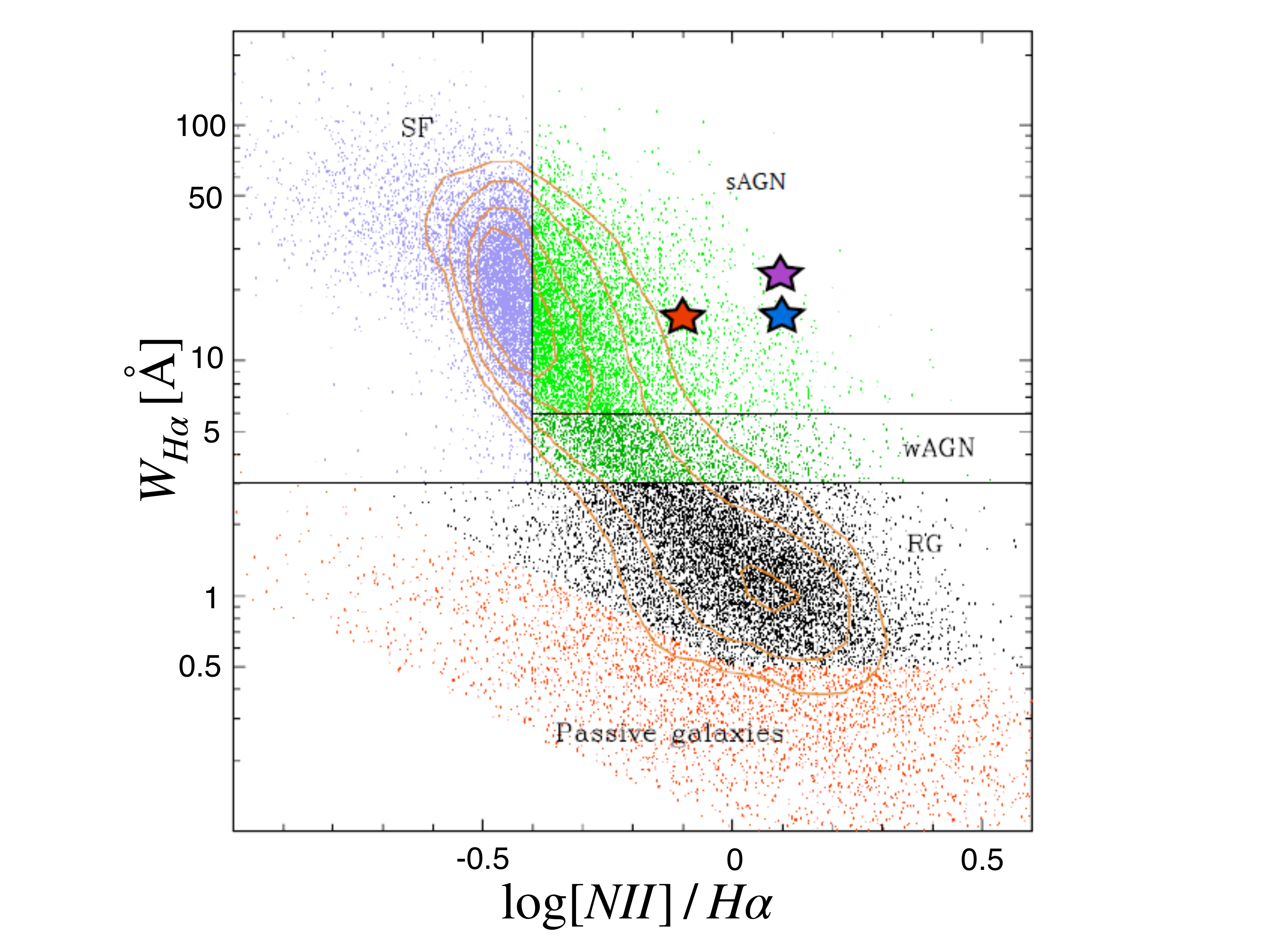}
\caption{The location in the WHAN diagram of three components of Mrk\,622 is shown: central (purple star), blue (blue star) and red (red star). This diagram was presented by \citet{2011MNRAS.413.1687C} and shows the distribution of 700,000 galaxies from the seven data-release of the SDSS. Their Figure\,6 distinguishes between strong AGN (sAGN), weak AGN (wAGN), retired galaxies (RG), passive galaxies and SF galaxies.}
\label{fig:Fig5}
\end{center}
\end{figure}

The case of Mrk\,622 is intriguing given the variations of its continuum and NLR line intensities, which suggest a variability time-span as short as 13 years (see Figure~\ref{fig:Fig1}). In general, the NLR is expected to vary on a much longer timescales (beyond several decades or more) as a result of a significant light-crossing timescale as well as longer recombination timescales. To qualify whether variations on a time-span less than a decade are real, an estimation of the NLR size is required in order to estimate its light-crossing time.

For this purpose, we adopted the single-zone approximation NLR model of \citet[][hereafter, BL05]{2005MNRAS.358.1043B}, who derived the relationship R$_{NLR}$\,=\,40\,L$^{0.45}_{44}$\,pc, where L$_{44}$ is the H$\beta$ luminosity in units of 10$^{44}$\,erg\,s$^{-1}$. The BL05 one-zone model assumes that H$\beta$, [OIII]$\lambda$5007 and [OIII]$\lambda$4363 originate in a single zone with homogeneous gas properties. Also, these authors present an approximation known as the two-zone model, in which the [OIII]$\lambda$5007 is assumed to be emitted in the outer region of the NLR where the density is lower, whereas the [OIII]$\lambda$4363 is produced in an inner high-density component.

The total estimated H$\beta$ luminosity (the luminosity from the three components) that was obtained from the \textit{WHT} spectrum is L$_{H\beta}$\,=\,2.41$\,\times\,$10$^{40}$\,$\pm$\,6.25\,$\,\times\,$\,10$^{37}$\,erg\,s$^{-1}$. Nevertheless, the Balmer decrement estimated also for the three components is H$\alpha$/H$\beta$\,=\,5.65$\pm$0.10, so it is convenient to make a correction due to intrinsic reddening. After applying this correction, and using the BL05 relationship, the size of the radius of the NLR is $R_{NLR}$\,=\,2.7\,pc, and the light-crossing time is $l_{ct}$\,=\,($R_{NLR})/c$\,=\,8.7\,yr. At face values, these estimates are consistent with the evidence of intrinsic variability of the lines, although we cannot rule out the possibility that the different pointing position of both telescopes might be the cause of the reported variations, given that the \textit{WHT} might potentially be sampling a more inner region of the NLR. 

\begin{figure*}
\includegraphics[width=1.5\columnwidth, height=7cm]{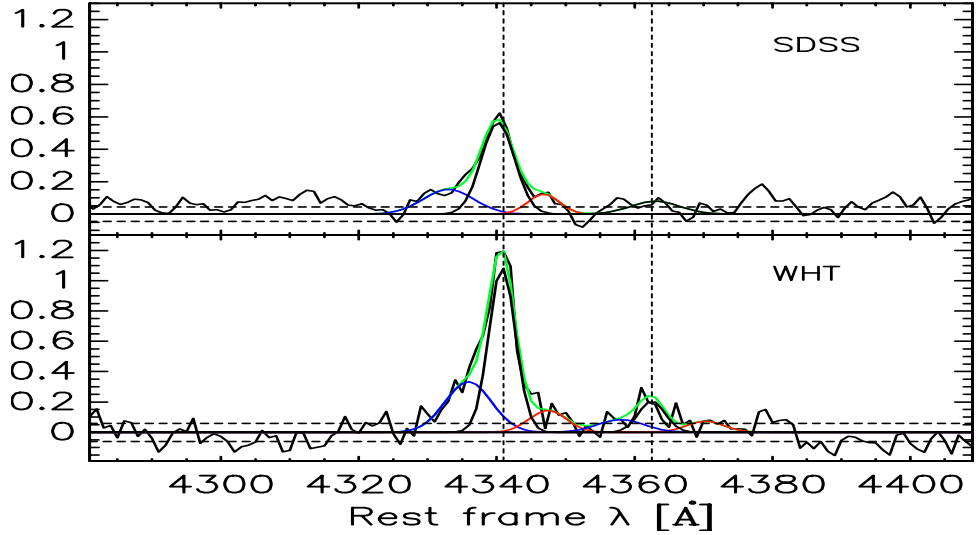}
\caption{3G model results in the region of H$\gamma\lambda$4341.  Upper panel shows the \textit{SDSS} spectrum. Lower panel the \textit{WHT} spectrum along with the detection of the [OIII]$\lambda$4363 emission line, that is find to be an upper limit. The blue and red components of the detected [OIII]$\lambda$\,4363 emission line are shown just for illustrative purposes, they have very poor S/N ratio, so only the central component was used to estimate the density of the NLR, see text.}
\label{fig:Fig6}
\end{figure*}

The [OIII]$\lambda$4363 emission line was detected with  the \textit{WHT} telescope, see the lower panel of Figure~\ref{fig:Fig6}. The blue and red components of this line are buried in the noise. Only the central component has enough signal-to-noise ratio (S/N) to enable the calculation of electron density in the framework of the BL05 one-zone model by means of the following ratios: $\log$([OIII]$\lambda$4363/[OIII]$\lambda$5007)\,=\,-0.8 and $\log$([OIII]$\lambda$5007/H$\beta$)\,=\,-0.2. This ratios were transformed to obtain the corresponding values of density $n_{e}$, see Figure 3 in BL05.
The results is $n_{e}\sim 3.5 \times 10^{6}$\,\cmc. Since the critical densities for [OIII]$\lambda$5007 and [OIII]$\lambda$4363 are $7\times10^{5}$\,\cmc\ and $3.3 \times 10^{7}$\,\cmc, respectively, the density found with our data and the one-zone BL05 approximation lies in between both critical density limits. The use of the [OIII]$\lambda$4363/[OIII]$\lambda$5007 ratio to estimate the density is valid when the density is not much above or below the two critical densities \citep{1989agna.book.....O}.

This result is consistent with our observations since it is compatible with NLR variations over a decade-long time-span. The other important timescale is that required for the ionised gas to reach a new equilibrium when the ionising source varies. The recombination timescale given by  
$\tau_{rec}$\,=\, ($n_{e}\alpha_{B}$)$^{-1}$\,s\, can be estimated using the derived electron density $n_e$ and assuming a T\,=\,10000$^{\circ}$K and case B recombination \citep[][]{2006agna.book.....O}. Therefore, for $n_{e}\,\sim\,3.5 \times 10^{6}$\,\cmc\ and a recombination coefficient of $\alpha_{B}\,= \,2.6\,\times\,10^{-13}$\,\cmcs\,, the recombination timescale is $\tau_{rec}$=\,1.1$\times10^{6}$\,s or $\sim$\,13 days. This result yields a very short recombination timescale. Therefore we can expect the [OIII] emission gas to rapidly change when the ionising continuum varies.

Using the \textit{WHT} spectrum, an estimate of the bulge stellar velocity dispersion and black hole mass for Mrk\,622 were given in Paper\,I, the obtained values are:  $\sigma_{\star}$\,=\,178\,$\pm$\,9\,km\,s$^{-1}$ and 
the log\,$M_{BH}$\,=\,7.50\,$\pm$\,0.12 in $M_{\sun}$ units. Using these estimates, an Eddington-ratio of $\lambda_{Edd}\sim$\,0.01 is obtained (see Table~\ref{table:Table4}). This is consistent with the derived value of the accretion rate $\dot{M}\sim\,$0.01\,$M_\odot$\,yr$^{-1}$. 
The radius of influence is $r_{BH}\,=\,\frac{G M_{BH}}{\sigma_{\star}^{2}}\,=\,$4.3$\pm$0.1\,pc, and hence is larger than the estimated NLR radius of 2.7\,pc. 

The density inferred using the one-zone model is worth noting as it reverses the trend found by \citet[][and references therein]{2001PASJ...53..629N, 2006A&A...459...55B} who showed convincing evidence that Sy\,1 NLR densities are much higher than that of Sy\,2's, yet Mrk\,622 belongs to Type\,II category since it has no visible\footnote{Alternatively, it is a CL AGN or its inner nucleus and accretion disk has recently turned off.} BLR. In short, the NLR densities of Mrk\,622 are high as far as the high excitation [OIII] emission gas is concerned and conversely occupy a small volume along the ionising cone axis. This factor would explain why the 4363\,\AA\ [OIII] line could vary on a relatively short timescale. To account for variations, we favor a model where the stronger nuclear ionising radiation reaching the NLR is much intense than 13 years ago, resulting in an increase of the 4363/5007 ratio. Such an ionisation front should lead to an increase in temperature until the photoionised gas locally reaches its new ionisation and thermal equilibrium. Time-dependent photoionisation calculations are in order.

In the context of LINERs, \citet{1995ApJ...445L...1E} calculated the evolution of the line intensities when the ionising source undergoes periodic pulse-like intensity variations. The authors averaged the line emission variations over many UV pulses and across a large NLR volume assumed. In this situation, light travel effects can be neglected since they cancel out. If adapted to Mrk\,622, such a scenario would require significant modifications. The time-dependent calculations shown in their Figure\,2, with $U=0.0045$ and a power-law $\propto \nu^{-1.5}$ with an on-pulse duration of $30$\% of the recombination timescale, show a decrease of a factor 2.7 of  4363/5007 from the peak value at the onset of the pulse with respect to the ratio near equilibrium value when the pulse turns off. Hence, an increase in intensity is expected to favor larger intensities for 4363 at the onset of the ionisation front.
 
The results from our analysis support the idea that the  intensity variations observed in the optical spectra of Mrk\,622, are produced inside a zone of the NLR that is very compact, has a very high density and lies close to the nucleus, favoring a time-dependent evolution of the line luminosities.
It is worth noting that  \citet{2014ApJ...796..134D} report NLR variability in the CL Sy Mrk\,590, where the distance of the NLR emission region is found to be $\lesssim$3\,pc\, from the central source.

\begin{table}
\caption{3G model results (\textit{WHT}). }
\label{table:Table1}
\begin{scriptsize}
\begin{tabular}{lcrrr}
\hline
\hline \\
Line$^{a}$ & $\lambda_{obs}$  & Flux$^{b}$~~  &  FWHM  & $EW$~~ \\
Id.  &  (\AA) & ($\,\times\,$10$^{-15}$)~~& (km\,s$^{-1}$) &  (\AA)~~   \\
\hline
\hb C & 4861.8\,$\pm$\,0.1 & 10.25\,$\pm$\,0.05 & 234\,$\pm$\,3 & 10.8\,$\pm$\,0.6 \\ 
\hb B & 4858.0\,$\pm$\,0.1 & 5.56\,$\pm$\,1.07 & 601\,$\pm$\,56 & 5.9\,$\pm$\,1.5 \\ 
\hb R & 4865.3\,$\pm$\,0.5 & 2.75\,$\pm$\,0.14 & 427\,$\pm$\,6 & 2.9\,$\pm$\,0.2 \\ \hline
\oiiia C & 4959.0\,$\pm$\,0.1 & 2.21\,$\pm$\,0.41 & 234\,$\pm$\,3 & 2.1\,$\pm$\,0.3 \\ 
\oiiia B & 4953.8\,$\pm$\,0.3 & 10.75\,$\pm$\,1.17 & 601\,$\pm$\,56 & 10.3\,$\pm$\,1.7 \\ 
\oiiia R & 4965.5\,$\pm$\,0.1 & 7.68\,$\pm$\,0.14 & 427\,$\pm$\,6 & 7.3\,$\pm$\,0.4 \\ \hline
\oiiib C & 5007.0\,$\pm$\,0.1 & 6.63\,$\pm$\,1.24 & 234\,$\pm$\,3 & 6.0\,$\pm$\,0.8 \\ 
\oiiib B & 5001.7\,$\pm$\,0.3 & 32.26\,$\pm$\,3.52 & 601\,$\pm$\,56 & 29.4\,$\pm$\,4.8 \\ 
\oiiib R & 5013.5\,$\pm$\,0.1 & 23.05\,$\pm$\,0.41 & 427\,$\pm$\,6 & 20.8\,$\pm$\,1.2 \\ \hline
\oia C & 6301.0\,$\pm$\,0.1 & 2.84\,$\pm$\,0.33 & 210\,$\pm$\,12 & 2.7\,$\pm$\,0.2 \\ 
\oia B & 6297.7\,$\pm$\,0.5 & 4.04\,$\pm$\,0.73 & 528\,$\pm$\,96 & 3.8\,$\pm$\,0.8 \\ 
\oia R & 6309.6\,$\pm$\,0.9 & 1.82\,$\pm$\,1.11 & 603\,$\pm$\,160 & 1.8\,$\pm$\,1.1 \\ \hline
\oib C & 6365.0\,$\pm$\,0.1 & 0.95\,$\pm$\,0.11 & 210\,$\pm$\,12 & 1.1\,$\pm$\,0.1 \\ 
\oib B & 6361.7\,$\pm$\,0.5 & 1.35\,$\pm$\,0.24 & 528\,$\pm$\,96 & 1.5\,$\pm$\,0.3 \\ 
\oib R & 6373.7\,$\pm$\,0.9 & 0.61\,$\pm$\,0.37 & 603\,$\pm$\,160 & 0.7\,$\pm$\,0.5 \\ \hline
\ha C & 6563.3\,$\pm$\,0.1 & 46.66\,$\pm$\,1.22 & 210\,$\pm$\,4 & 44.7\,$\pm$\,2.1 \\ 
\ha B & 6558.4\,$\pm$\,0.9 & 31.05\,$\pm$\,2.76 & 528\,$\pm$\,68 & 29.8\,$\pm$\,2.7 \\ 
\ha R & 6567.8\,$\pm$\,0.5 & 27.15\,$\pm$\,2.43 & 603\,$\pm$\,96 & 25.9\,$\pm$\,2.2 \\ \hline
\niia C & 6548.4\,$\pm$\,0.1 & 12.31\,$\pm$\,0.45 & 210\,$\pm$\,4 & 11.9\,$\pm$\,0.5 \\ 
\niia B & 6543.6\,$\pm$\,0.6 & 12.97\,$\pm$\,0.98 & 528\,$\pm$\,68 & 12.6\,$\pm$\,1.0 \\ 
\niia R & 6552.1\,$\pm$\,0.5 & 8.46\,$\pm$\,0.52 & 603\,$\pm$\,96 & 8.2\,$\pm$\,0.8 \\ \hline
\niib C & 6583.9\,$\pm$\,0.1 & 36.98\,$\pm$\,1.34 & 210\,$\pm$\,4 & 34.8\,$\pm$\,1.4 \\ 
\niib B & 6580.0\,$\pm$\,0.5 & 38.94\,$\pm$\,2.94 & 528\,$\pm$\,68 & 36.8\,$\pm$\,2.8 \\ 
\niib R & 6589.7\,$\pm$\,0.5 & 25.40\,$\pm$\,1.55 & 603\,$\pm$\,96 & 23.8\,$\pm$\,2.5 \\ \hline
\siia C & 6717.2\,$\pm$\,0.1 & 7.50\,$\pm$\,0.73 & 210\,$\pm$\,4 & 6.6\,$\pm$\,0.6 \\ 
\siia B & 6712.3\,$\pm$\,1.0 & 5.82\,$\pm$\,0.63 & 528\,$\pm$\,68 & 5.1\,$\pm$\,0.6 \\ 
\siia R & 6721.7\,$\pm$\,0.2 & 4.21\,$\pm$\,0.93 & 603\,$\pm$\,96 & 3.7\,$\pm$\,1.0 \\ \hline
\siib C & 6731.3\,$\pm$\,0.1 & 7.80\,$\pm$\,0.55 & 210\,$\pm$\,4 & 7.1\,$\pm$\,0.5 \\ 
\siib B & 6723.3\,$\pm$\,0.1 & 4.46\,$\pm$\,0.51 & 528\,$\pm$\,68 & 4.0\,$\pm$\,0.6 \\ 
\siib R & 6735.0\,$\pm$\,1.2 & 6.19\,$\pm$\,1.17 & 603\,$\pm$\,96 & 5.6\,$\pm$\,1.1 \\ 
\hline \hline
\multicolumn{5}{l}{$^{a}$ Stands for central(C),  blue (B) and red (R) components.} \\
\multicolumn{5}{l}{$^{b}$ Fluxes units: erg\,s$^{-1}$\,cm$^{-2}$\,\AA$^{-1}$.} \\
\end{tabular}
\end{scriptsize}
\end{table}

\begin{table}
\caption{3G model results (\textit{SDSS})}
\label{table:Table2}
\label{parsdss}
\begin{scriptsize}
\begin{tabular}{lcrrr}
\hline
\hline \\
Line$^{a}$ & $\lambda_{obs}$  & Flux$^{b}$~~  &  FWHM  & $EW$~~ \\
Id.  &  (\AA) & ($\,\times\,$10$^{-15}$)~~    & (km\,s$^{-1}$) &  (\AA)~~   \\
\hline
\hb C & 4861.1\,$\pm$\,0.1 & 6.97\,$\pm$\,1.75 & 314\,$\pm$\,25 & 7.5\,$\pm$\,1.8 \\ 
\hb B & 4856.7\,$\pm$\,1.2 & 3.03\,$\pm$\,1.14 & 619\,$\pm$\,91 & 3.2\,$\pm$\,1.3 \\ 
\hb R & 4864.5\,$\pm$\,1.3 & 2.71\,$\pm$\,0.99 & 523\,$\pm$\,54 & 2.9\,$\pm$\,1.1 \\ 
\hline
\oiiia C & 4958.0\,$\pm$\,0.5 & 1.10\,$\pm$\,0.65 & 314\,$\pm$\,25 & 1.1\,$\pm$\,0.7 \\ 
\oiiia B & 4952.5\,$\pm$\,0.7 & 5.69\,$\pm$\,0.97 & 619\,$\pm$\,91 & 5.9\,$\pm$\,1.1 \\ 
\oiiia R & 4964.4\,$\pm$\,0.1 & 4.79\,$\pm$\,0.45 & 523\,$\pm$\,54 & 5.0\,$\pm$\,0.7 \\ 
\hline
\oiiib C & 5006.0\,$\pm$\,0.5 & 3.29\,$\pm$\,1.94 & 314\,$\pm$\,25 & 3.4\,$\pm$\,2.0 \\ 
\oiiib B & 5000.5\,$\pm$\,0.7 & 17.08\,$\pm$\,2.91 & 619\,$\pm$\,91 & 17.6\,$\pm$\,1.7 \\ 
\oiiib R & 5012.5\,$\pm$\,0.1 & 14.38\,$\pm$\,1.36 & 523\,$\pm$\,54 & 14.7\,$\pm$\,2.2 \\ 
\hline
\oia C & 6300.8\,$\pm$\,0.5 & 1.41\,$\pm$\,0.62 & 276\,$\pm$\,91 & 1.4\,$\pm$\,1.1 \\ 
\oia B & 6297.6\,$\pm$\,2.8 & 1.83\,$\pm$\,0.84 & 513\,$\pm$\,54 & 1.8\,$\pm$\,1.0 \\ 
\oia R & 6305.1\,$\pm$\,1.8 & 1.82\,$\pm$\,0.77 & 548\,$\pm$\,25 & 1.8\,$\pm$\,1.2 \\ 
\hline
\oib C & 6364.9\,$\pm$\,0.5 & 0.47\,$\pm$\,0.37 & 276\,$\pm$\,91 & 0.5\,$\pm$\,0.4 \\ 
\oib B & 6361.5\,$\pm$\,2.4 & 0.61\,$\pm$\,0.34 & 513\,$\pm$\,54 & 0.6\,$\pm$\,0.4 \\ 
\oib R & 6369.1\,$\pm$\,1.8 & 0.61\,$\pm$\,0.30 & 548\,$\pm$\,25 & 0.6\,$\pm$\,0.4 \\ 
\hline
\ha C & 6562.8\,$\pm$\,0.1 & 25.65\,$\pm$\,4.62 & 276\,$\pm$\,22 & 24.5\,$\pm$\,4.6 \\ 
\ha B & 6557.3\,$\pm$\,0.6 & 17.40\,$\pm$\,2.46 & 513\,$\pm$\,68 & 16.7\,$\pm$\,2.5 \\ 
\ha R & 6566.6\,$\pm$\,0.9 & 16.47\,$\pm$\,3.27 & 548\,$\pm$\,67 & 15.8\,$\pm$\,2.0 \\ 
\hline
\niia C & 6548.0\,$\pm$\,0.3 & 7.09\,$\pm$\,0.64 & 276\,$\pm$\,22 & 6.8\,$\pm$\,0.7 \\ 
\niia B & 6542.7\,$\pm$\,0.6 & 6.96\,$\pm$\,0.98 & 513\,$\pm$\,68 & 6.7\,$\pm$\,0.9 \\ 
\niia R & 6548.6\,$\pm$\,0.9 & 4.72\,$\pm$\,0.36 & 548\,$\pm$\,67 & 4.5\,$\pm$\,0.4 \\ 
\hline
\niib C & 6583.4\,$\pm$\,0.1 & 21.28\,$\pm$\,1.92 & 276\,$\pm$\,22 & 20.3\,$\pm$\,2.0 \\ 
\niib B & 6578.9\,$\pm$\,0.7 & 20.90\,$\pm$\,2.95 & 513\,$\pm$\,68 & 20.0\,$\pm$\,2.7 \\ 
\niib R & 6588.8\,$\pm$\,0.6 & 14.17\,$\pm$\,1.09 & 548\,$\pm$\,67 & 13.5\,$\pm$\,1.2 \\ 
\hline
\siia C & 6716.9\,$\pm$\,0.1 & 5.99\,$\pm$\,0.46 & 276\,$\pm$\,22 & 5.7\,$\pm$\,0.6 \\ 
\siia B & 6712.3\,$\pm$\,1.1 & 3.95\,$\pm$\,0.33 & 513\,$\pm$\,68 & 3.7\,$\pm$\,0.4 \\ 
\siia R & 6721.3\,$\pm$\,1.0 & 2.99\,$\pm$\,0.49 & 548\,$\pm$\,67 & 2.8\,$\pm$\,0.6 \\ 
\hline
\siib C & 6731.1\,$\pm$\,0.1 & 5.19\,$\pm$\,0.46 & 276\,$\pm$\,22 & 4.9\,$\pm$\,0.5 \\ 
\siib B & 6726.8\,$\pm$\,0.7 & 3.10\,$\pm$\,0.77 & 513\,$\pm$\,68 & 2.9\,$\pm$\,0.8 \\ 
\siib R & 6734.9\,$\pm$\,2.0 & 2.83\,$\pm$\,0.73 & 548\,$\pm$\,67 & 2.7\,$\pm$\,0.7 \\ 
\hline
\hline
\multicolumn{5}{l}{$^{a}$ Stands for central(C),  blue (B) and red (R) components.} \\
\multicolumn{5}{l}{$^{b}$ Fluxes units: erg\,s$^{-1}$\,cm$^{-2}$\,\AA$^{-1}$.} \\
\end{tabular}
\end{scriptsize}
\end{table}


\begin{table}
\caption{3G model results for H$\gamma$ and [OIII]$\lambda$4363}
\label{table:Table3}
\begin{scriptsize}
\begin{tabular}{lcrrr}
\hline
\hline \\
Line$^{a}$ & $\lambda_{obs}$  & Flux$^{b}$~~  &  FWHM  & $EW$~~ \\
Id.  &  (\AA) & ($\,\times\,$10$^{-15}$)~~    & (km\,s$^{-1}$) &  (\AA)~~   \\
\hline
\textit{WHT}\\
\hg C	&	4340.8	\,$\pm$\,	0.2	&	5.25	\,$\pm$\,	1.35	&	313	\,$\pm$\,	52	&	5.6	\,$\pm$\,	1.4	\\
\hg B	&	4335.9	\,$\pm$\,	1.5	&	2.79	\,$\pm$\,	1.12	&	545	\,$\pm$\,	150	&	3.0	\,$\pm$\,	1.2	\\
\hg R	&	4347.4	\,$\pm$\,	1.4	&	0.98	\,$\pm$\,	0.28	&	441	\,$\pm$\,	167	&	1.0	\,$\pm$\,	0.4	\\
\oiii{$\lambda$4363} C	&	4362.5	\,$\pm$\,	1.0	&	1.05	\,$\pm$\,	0.85	&	313	\,$\pm$\,	52	&	1.0	\,$\pm$\,	1.1	\\
\oiii{$\lambda$4363} B	&	4358.0	\,$\pm$\,	1.2	&	0.71	\,$\pm$\,	0.70	&	545	\,$\pm$\,	150	&	0.7	\,$\pm$\,	0.8	\\
\oiii{$\lambda$4363} R	&	4370.3	\,$\pm$\,	1.4	&	0.50	\,$\pm$\,	0.29	&	441	\,$\pm$\,	167	&	0.5	\,$\pm$\,	0.4	\\
\hline
\textit{SDSS}\\
\hg C &4340.2\,$\pm$\,0.4&3.40\,$\pm$\,0.59&390\,$\pm$\,46&3.6\,$\pm$\,0.7\\
\hg B &4333.0\,$\pm$\,2.4&1.40\,$\pm$\,0.81&600\,$\pm$\,261&1.5\,$\pm$\,0.9\\
\hg R & 4346.8\,$\pm$\,1.2 & 0.70\,$\pm$\,0.25 & 370\,$\pm$\,163 & 0.7\,$\pm$\,0.3\\
\oiii{4363} C & 4363.0\,$\pm$\,0.7 & 0.70\,$\pm$\,0.12 & 570\,$\pm$\,202  & 0.7\,$\pm$\,0.1\\
\hline
\hline
\multicolumn{5}{l}{$^{a}$ Stands for central(C),  blue (B) and red (R) components.} \\
\multicolumn{5}{l}{$^{b}$ Fluxes units: erg\,s$^{-1}$\,cm$^{-2}$\,\AA$^{-1}$.} \\
\end{tabular}
\end{scriptsize}
\end{table}

\begin{table}
\caption{Eddington and bolometric luminosities,  
Eddington ratio 
and mass accretion rate 
for Mrk\,622}
\label{table:Table4}
\begin{tabular}{lc}
\hline
L$_{Edd}$ = 39.78\,$\pm$\,0.63 $^{(a)}$ \\

L$_{bol}$ \,=\, 600\,L$_{\lambda 5007}$ \,=\, 0.46\,$\pm$\,0.05\,$^{(b)}$\\
$\lambda_{Edd}$= L$_{bol}$/L$_{Edd}$ \,=\, 0.012\,$\pm$\,0.001 $^{(c)}$\\	
~~$\dot{M}$~~\,=\,L$_{bol}$/$\eta$\,c$^{2}$ 
\,=\, 8\,$\,\times\,$\,10$^{-3}\,\pm$\, 9\,$\,\times\,$\,10$^{-4}$ M$_{\odot}$\,yr$^{-1}$ $^{(d)}$ \\
\hline 
\multicolumn{1}{l}{$^{(a)}$ In units of  10$^{44}$\,erg\,s$^{-1}$ }\\ 
\multicolumn{1}{l}{$^{(b)}$ In units of  10$^{44}$\,erg\,s$^{-1}$ 
\citep[see][]{2009MNRAS.397..135K}.}\\
\multicolumn{1}{l}{$^{(c)}$ $\lambda_{Edd}$ is the Eddington ratio \citep[see][]{2008ARA..46..475H}. } \\
\multicolumn{1}{l}{$^{(d)}$ $\dot{M}$ is the mass accretion rate, estimated with $\eta$=0.1.  } 
\end{tabular}
\end{table}

\section{NIR Photometric Decomposition}
\label{phot}

\begin{figure*}
\includegraphics[scale=0.4]{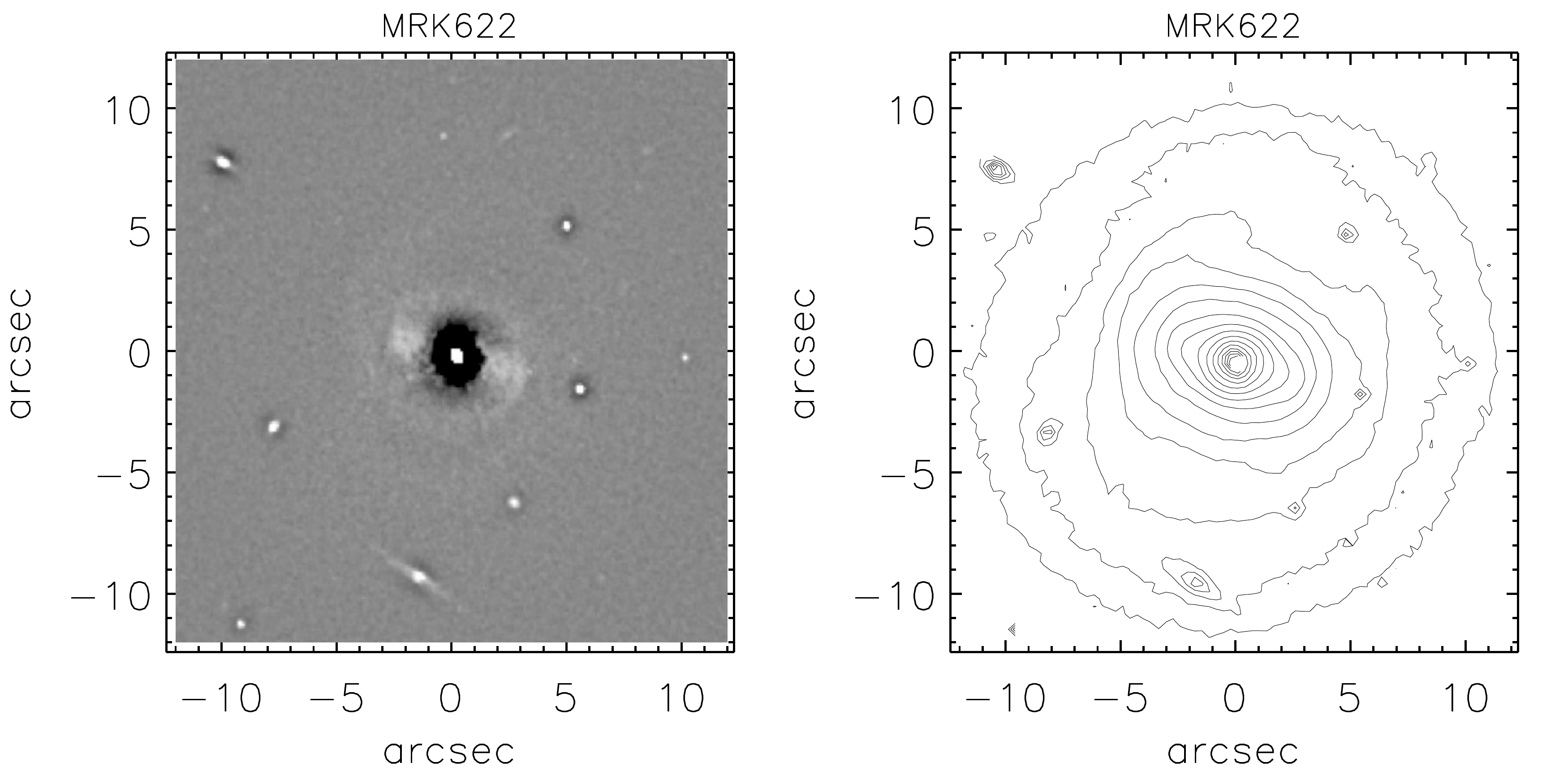}
\caption{Left panel: unsharp mask of a region of size 250 pixels or (20$\,\times\,$20\arcsec) done within a radius of 3 pixels. The bar is clearly shown, as well as a trace of a faint spiral structure. Right panel: isophotal map of the same region showing a barred structure.
}
\label{fig:Fig7}
\end{figure*}

An HST archival image was downloaded and analysed in order to study the structural parameters of the host galaxy of Mrk\,622. This image was acquired with the \textit{WFC3} in March 12 2012  with the F105W filter (Y-wide, central $\lambda\,=\,$1055.2 nm and width\,=\,265 nm), with an exposure time of 238.7 seconds (proposal 12521 PI: X. Liu). Taking a look at the image it is possible to see the existence of a bar in the host galaxy. The left panel of {Figure~\ref{fig:Fig7}} shows the result obtained after unsharp masking the image. The right panel shows logarithmic contours.  Both panels clearly show the presence of a bar-like structure. Following the method described by \citet{2004A&A...415..941E} and references therein, it is found that the bar has a semi-major axis \rm{L}$_{bar}$\,=\,3.85\arcsec\,$\pm$\,0.26\arcsec, adopting as \rm{L}$_{bar}$ the position where the ellipticity is largest.  At the object's distance, this size corresponds to 1.82\,kpc, so the length of the bar is 3.64\,kpc. The bar has a position angle of $\sim$74.0\,$\pm$\,1.5 degrees and it is somewhat a small bar for an early-type galaxy, but larger than a typical secondary or inner bar, which usually extends from 240 to 750\,pc \citep[see][]{2002AJ....124...65E}. 

On the other hand, the HST NIR image was modelled using Imfit \citep{2015ApJ...799..226E}\footnote{Imfit: http://www.mpe.mpg.de/~erwin/index.html} with a S\'ersic profile for the bar and bulge and an exponential law for the disk \citep[][]{1968adga.book.....S}. The results show that the bulge is best fitted with a S{\'e}rsic index $n\,=\,$1.128\,$\pm$\,0.004. The bar was best fitted using boxy ellipses with a S\'ersic index $n$\,=\,0.605\,$\pm$\,0.001 and $PA$\,=\,74.6\,$\pm$\,0.02 (in agreement with the value calculated with the method described in the previous paragraph). 

A spheroidal component (a classical a bulge) is proposed to form as a result of rich galaxy mergers \citep[e.g.,][]{1991ApJ...370L..65B,2010ApJ...715..202H}. Classical bulges have a S\'ersic index $n$ between 2\,$\leq\,$n$\,\leq$4. On the other hand, the so-called pseudo\-bulges are found to be formed through dynamical instabilities in disk galaxies. Angular momentum transport by the bar builds inner and outer rings and drives gas to the center where SF is triggered.  They are dynamically cold and have S\'ersic indices $n\,\leq$\,2, with velocity dispersion intermediate between those of classical bulges and disks \citep[see][]{2004ARA&A..42..603K,2008AJ....136..773F}.

\begin{figure*}
\includegraphics[width=\textwidth]{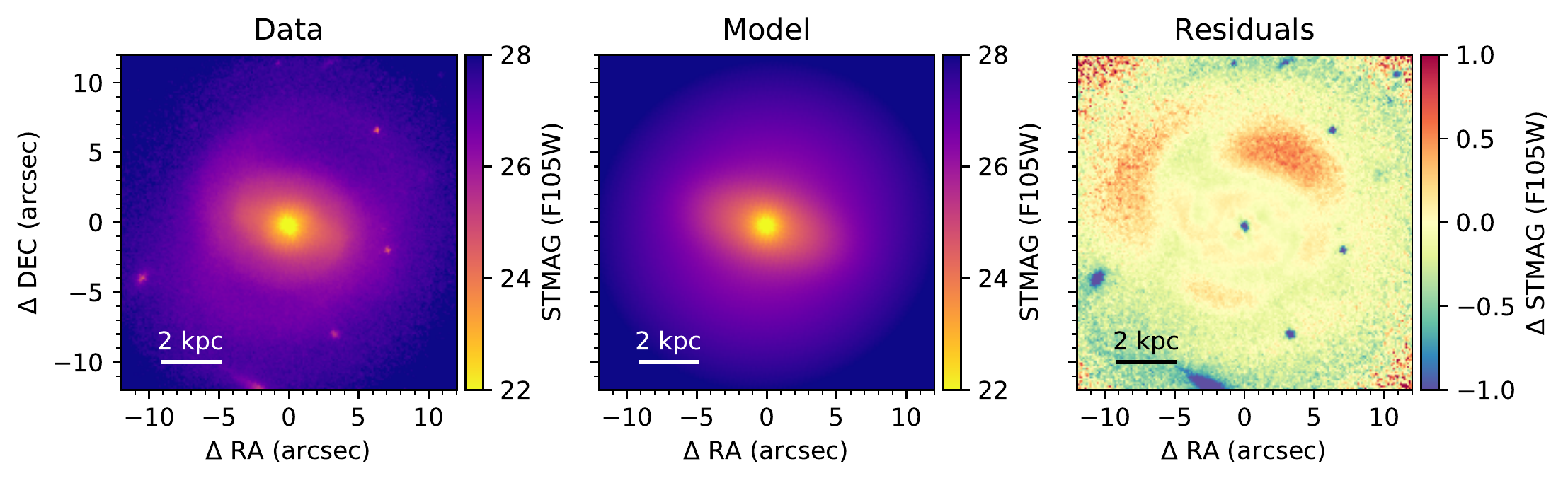}
\caption{Imfit results of the modelling of the NIR HST image of Mrk\,622. The residuals image shows a relaxed system without signs of interactions. }
\label{fig:Fig8}
\end{figure*}

Therefore, the structural parameters found in Mrk\,622 hint towards secular evolution of the host galaxy \citep[see][and references therein]{2013seg..book....1K}. This is supported by its morphology; the galaxy shows an inner ring, that probably coincides with the inner Lindblad resonance (ILR) and also an outer and fainter external ring. So, the de Vaucouleurs classification for the host galaxy is (R)SB(rs)b. The left panel of {Figure~\ref{fig:Fig8}} shows the HST archival NIR image, middle shows the best model obtained, and the right panel shows the residual image. The residual image shows no evidences of tidal tails. So, if a merger did happened it is now in a late evolutionary phase. The bar found can drive cold gas to the nuclear region that accumulates and triggers star formation \citep[e.g.,][]{2007IAUS..235...19C}. 

The plate scale of the HST image is 60.8\,pc per pixel at the distance of Mrk\,622. Therefore, the NIR HST image cannot resolve the nuclear region at the resolution needed to show a possible double nucleus separated by $\sim$76\,pc\, (0.16\arcsec, see section~\ref{spectral} and
Figure~\ref{fig:multi}(b)).

\begin{figure*}
\includegraphics[width=2.0\columnwidth]{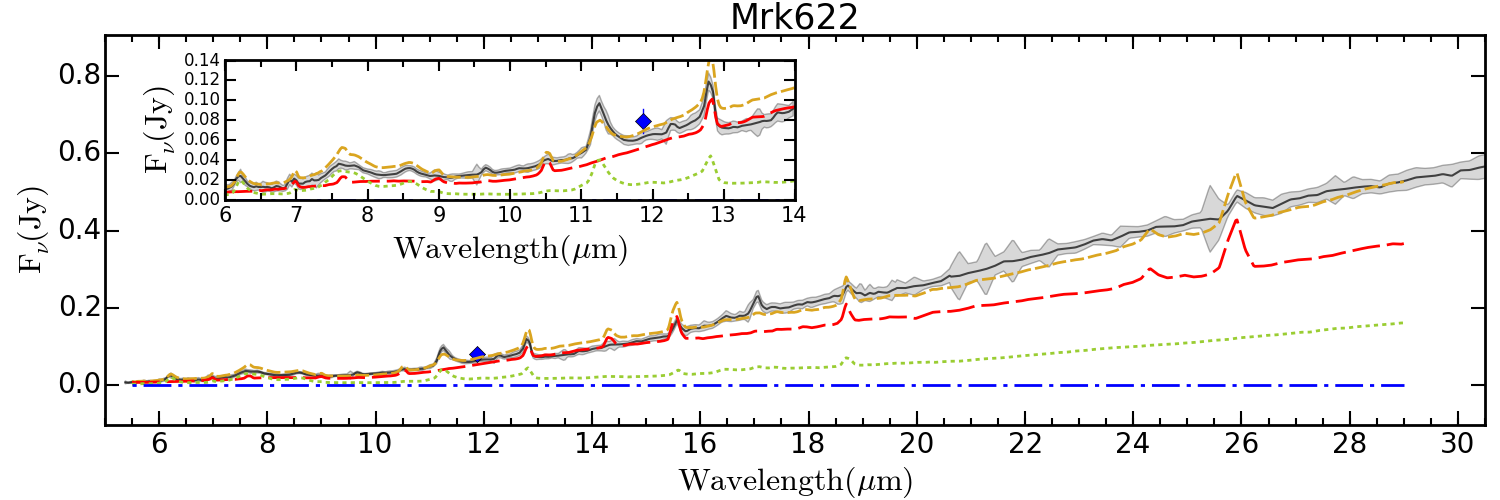}
\includegraphics[width=2.0\columnwidth]{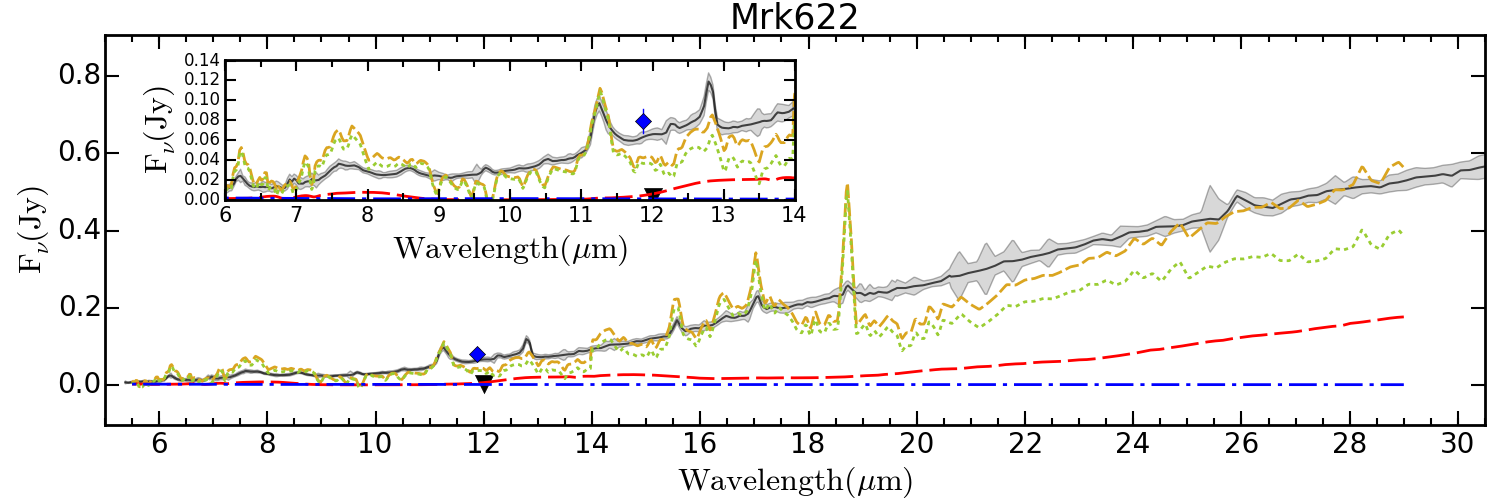}
\caption{\textit{IRS/\textit{Spitzer}} spectra of Mrk\,622 
(grey shaded area and the black continuous line), and its best fit before (top) and after (bottom) including the 2-10\,keV 
observed flux as a constraint to the AGN component (see text). The final best fit, 
AGN, stellar, and ISM components are shown as yellow short-dashed, red long-dashed, blue dot-dashed, and green dotted lines respectively. The high spatial resolution \textit{CanariCam} 11.6 $\rm{\mu m}$ flux is shown as a blue diamond. The expected  12 $\rm{\mu m}$ flux (obtained from the 2-10\,keV\, X-ray luminosity, see text) is shown as a black inverted triangle. The small inset panels show the 6 to 14 $\rm{\mu m}$ range.}
\label{fig:Fig9}
\end{figure*}

\section{Mid-infrared observations and data reduction}
\label{mir-obs}

High spatial resolution mid-infrared images of Mrk\,622 were obtained using \textit{CanariCam}  
\citep{2003SPIE.4841..913T} on the 10.4m \textit{Gran Telescopio Canarias} \textit{GTC} 
(TC3-14BIACMEX, P.I.: I. Cruz-Gonz\'alez). Observations were done in December 5 2014
using the Si5 filter at 11.6\,$\rm{\mu m}$. The on-source integration time was 1260 seconds. \textit{CanariCam} uses a Raytheon 320$\rm{\times}$240 Si:As detector that covers a field of view (FoV) of 26\arcsec$\rm{\times}$19\arcsec on the sky with a pixel scale of 0.08\arcsec. \textit{CanariCam} has a spatial resolution range of 0.3\arcsec to 0.6\arcsec, depending on wavelength, which corresponds to 142 up to 283\,pc\, at the distance of the object. Standard mid-infrared chopping-nodding techniques were used to remove the time-variable sky background, the thermal emission from the telescope, and the detector 1/f noise. The instrument position angle, chopping and nodding throws, chop and nod position angles were 90$\rm^{\circ}$, 10\arcsec, 10\arcsec, -180$\rm^{\circ}$, and 0$\rm^{\circ}$, respectively. An image of the point spread function (PSF) of the standard star was obtained with the same filter immediately before the science target, to accurately sample the image quality and for flux calibration of the target observation. The exposure time for this standard star was set to 66 seconds. 

Each observing block was processed using the pipeline \textit{RedCan} \citep{2013A&A...553A..35G}. This process enables to produce flux-calibrated images and wavelength- and flux-calibrated spectra for \textit{CanariCam/GTC} and \textit{T-ReCS/Gemini} low-resolution data. Using the standard star we have computed a flux calibration factor of 4.27$\rm{\times}10^{-5}$ Jy ADU$^{-1}$ $\rm{s^{-1}}$ (included automatically by \textit{RedCan} as a keyword called FLUXCAL in the headers).

\begin{table}
\caption{Mid-Infrared Observations with \textit{CanariCam/GTC}}
\label{table:Table5}
\begin{tabular}{ll}
\hline 
\hline \\
Parameter  & Value   \\
\hline
Integration time & 1260 s \\
Wavelength & 11.6\,$\mu$m \\
Observation date & 2014 Dec 5 \\
Total flux & 79$\pm$12 mJy ($\rm{2.0\,\pm\,0.3\,\times\,10^{-11} erg\,s^{-1}\,cm^{-2}}$)		\\
Size$^{a}$  & $<$0.40\arcsec\,$\,\times\,$\,0.34\arcsec$^{a}$ \\
& ($<$189\,pc\,$\,\times\,$\,175\,pc\, at D=101.9 Mpc) \\
\hline									
\hline \\
\multicolumn{2}{l}{$^{a}$ Unresolved.}
\end{tabular}
\end{table}

In order to investigate the extension of the target, the image of the standard star and the target were fitted with a 2D Gaussian profile. The PSF of the standard star has a width of 0.40\arcsec\,${\times}\,$0.34\arcsec, while the target width is 0.40\arcsec\,$\,\times\,$\,0.37\arcsec. So, the widths found for both the star and the source are quite similar. Therefore, we infer that the mid-infrared image of Mrk\,622 does not allow resolving the nuclear source (see Figure~\ref{fig:multi}(c)). At the distance of Mrk\,622 this width corresponds to 189\,pc$\rm{\,\times}$\,175\,pc. By fitting two 2D Gaussian profiles to the target source we find that any possible extended structure is negligible (contributing less than 2\% of the flux). Finally, the total flux of Mrk\,622 calculated at 11.6 $\rm{\mu m}$ is 79\,$\rm{\pm}$\,12 mJy, i.e., $\rm{(2.0\,\pm\,0.3)\,\times\,10^{-11}}$\, erg\,s$^{-1}$\,cm$^{-2}$. The \textit{CanariCam/GTC} observations are summarized in Table~\ref{table:Table4}.

Ground-based \textit{CanariCam}/GTC images were combined with archival \textit{Spitzer}/IRS low resolution spectrum downloaded from the Combined atlas of sources with \textit{Spitzer}/IRS spectra (CASSIS\footnote{http://cassis.sirtf.com}). The spectrum was observed on April 15 2005. CASSIS automatically identified the spectrum as point-like (i.e., 3.6\,--\,4.5\,\arcsec) and therefore used optimal extraction to produce the resulting spectrum. The code {\it decompIRS}  \citep{2015ApJ...803..109H} was used to decompose the \textit{Spitzer}/IRS spectra into the nuclear (i.e., AGN dust) and circumnuclear (ISM and stellar) components. 
The result of this decomposition is shown at the top panel of Figure~\ref{fig:Fig9}. 
According to this decomposition, above 90\% of the \textit{Spitzer}/IRS 12~$\rm{\mu m}$ flux is dominated by AGN dust. However, we note that this decomposition might not be correct according to the X-ray luminosity of Mrk\,622 (see Section\,\ref{xray}).

\section{MIR and X-ray data analysis}
\label{xray}

Mrk\,622 was observed by XMM-Newton in 2003. Based on this observation two independent analysis were performed by \citet{2005A&A...444..119G} and \citet{2014A&A...569A..71C}, henceforth G05 and C14 respectively, both of them as part of a systematic analysis of samples of obscured AGN. The S/N ratio of the data does not allow a very detailed analysis, however some important parameters can be re\-covered. According to G05, the source can be fitted with a composite model that includes a power-law associated to the AGN innermost emission, a thermal emission from a collisionally ionised plasma \citep{1985A&AS...62..197M}, and a narrow Gaussian Fe K$\alpha$ emission line at 6.4\,keV. The index of the power-law was fixed to 2, the typical value for AGN \citep[e.g.,][]{2005A&A...432..835P}. The temperature associated to the measured thermal emission is 0.59$^{+0.09}_{-0.21}$\,keV and the upper limit for the equivalent width ($EW$) of the iron line is 1600\,eV. The 2-10 keV and [OIII] flux ratio indicates a Compton-Thick (CT) nature, see G05. Therefore, the hydrogen column density ($N_H$) was assumed to be higher than 1.6\,$\times\,10^{24}$\,cm$^{-2}$. The high value of the $EW$ of the Fe K$_\alpha$ line is also an indication of the CT nature of Mrk\,622. Observations by G05 report a 2-10 keV flux of 2.2\,$\times\,10^{-13}$\,erg\,s$^{-1}$\,cm$^{-2}$.

A separate analysis of the same XMM-Newton observation was performed by C14. In this case, the best-fit model includes a double power-law model plus, a narrow emission Fe line and a thermal component: {\it zwabs(zpowerlaw +Feline)+ (zpowerlaw + mekal}). This model accounts for the intrinsic emission of the AGN (power-law and Fe emission line) which is partially absorbed and partially scattered by absorbing material in our line-of-sight. The thermal contribution is not affected by the absorbing material. The indices of the power-law components were fixed to 1.9 following the same argument of G05. The contribution of the scattered emission of the power-law is only 6\%. The resulting temperature of the thermal component is 0.7$\pm0.2$\,keV and the $EW$ of the iron line is 900$^{+700}_{-600}$\,eV. All of these parameters are compatible within the errors with the results obtained by G05. Interestingly, C14 left the value of the $N_H$ free, measuring a value for the $N_H$ of 4$^{+3}_{-2}\times10^{23}$\,cm$^{-2}$. Although the value of $N_H$ is well below the limit established for CT objects, that is $1.5\times10^{24}$\,cm$^{-2}$, the authors classify Mrk\,622 as a CT candidate, based in the high measured $EW$ of FeK$\alpha$ line. The observed flux and the intrinsic luminosity measured in the 2-10 KeV band by C14 are 2.15\,$\times\,10^{-13}$\,erg\,s$^{-1}$\,cm$^{-2}$ and 8.71\,$\times\,10^{41}$\,erg\,s$^{-1}$, respectively. 
 
Many authors have found a tight correlation between the 2-10 keV X-ray and the 12\,$\rm{\mu m}$ emissions \citep[e.g.,][]{2009A&A...502..457G,2015MNRAS.454..766A}. If the whole X-ray emission, in the 2-10\,keV band, is assumed to arise in the AGN, an estimation of the expected 12\,$\rm{\mu m}$ flux associated to the AGN can be obtained. This assumption is very reasonable as we expect a very minor contribution in the 2-10\,keV of X-ray flux associated to a thermal emission at the temperature obtained for Mrk\,622, 0.4-0.9\,keV. Therefore, using the correlation in \citet{2009A&A...502..457G}\footnote{$\log{F_{12\rm{\mu m}}}\,+\,11\,=\,0.35\,+\,0.79\,(\log{F_{2-10\,keV}}\,+\,11)$} and the C14 measurement of the flux $F_{2-10\,keV}$, the estimated flux at 12\,$\rm{\mu m}$ associated to the AGN is only 4.4\,mJy. This value corresponds to 6\% of the total flux measured by \textit{CanariCam}/GTC. 

In order to understand this discrepancy three scenarios have been considered: The first one takes into account that the contribution of the AGN represents only the 6\% of the whole MIR emission measured by \textit{CanariCam}/GTC and the remaining comes from other contributors. In order to test this scenario the \textit{Spitzer}/IRS spectrum of Mrk\,622 has been fitted using the {\it DecompIRS} code. 

The \textit{Spitzer}/IRS spectrum of Mrk\,622 was downloaded using CASSIS\footnote{http://cassis.sirtf.com}. The spectrum was extracted as point-like and the estimated limit on the size of the point-like source is 1.13\arcsec~  (i.e., $\rm{\sim}$545\,pc). Figure~\ref{fig:Fig9} (top) shows the \textit{Spitzer}/IRS  \citep{2015ApJ...803..109H} spectrum and the \textit{CanariCam} monochromatic flux at 11.6\,$\rm{\mu m}$ obtained in this analysis (blue diamond). 
Both spectrum and the photometric point are in good agreement, suggesting that both apertures are dominated by the same emission. The {\it DecompIRS} code  uses empirical spectra to decompose \textit{Spitzer}/IRS spectra into three components: AGN (red long-dashed line), stellar (blue dot-dashed line) and ISM (or PAH dominated, green dotted line), see top panel of Figure~\ref{fig:Fig9}. The yellow short-dashed shows the combined spectrum.  In spite of the strength of some emission lines, the combined spectrum reproduces well the observed \textit{Spitzer}/IRS spectrum. The AGN component dominates the emission, contributing in 75\,\% to the total emission in the 5-29\,$\rm{\mu m}$ region. The stellar component is negligible and the ISM component contributes in 25\,\% to the 5-29\,$\rm{\mu m}$ emission. The flux associated to the AGN component at 12\,$\rm{\mu m}$ is $\rm{\sim 58}$ mJy. According to the results of this fit, the most important contribution comes from the AGN, followed by the ISM, and a minor contribution of the stellar population is reported. This result does not agree with having a 6\,\% contribution of the AGN to the \textit{CanariCam}/GTC fluxed derived from the X-ray flux estimation. 

The \textit{Spitzer}/IRS spectrum was fitted again but now fixing the contribution of the AGN at 12\,$\rm{\mu m}$ to the value derived from the X-ray flux of 4.4\,mJy. The contribution of the stellar and ISM component were then left free to vary to fit the data. The result of this modeled spectrum can be seen in the lower panel of Figure~\ref{fig:Fig9}. In this model, the most important contribution to the IR emission comes from the ISM. Although a reasonable fit is found below 17\,$\rm{\mu m}$, the model fits poorly the data between 17 and 25\,$\rm{\mu m}$. Therefore, we can conclude that a major contribution of the ISM or stellar components to the whole IR emission to explain the discrepancy between X-ray and IR flux is not plausible.

The possibility that the absorption of Mrk\,622 is underestimated in the X-ray analysis performed by C14 is the third scenario considered. The discrepancy between X-ray and MIR emission emerges when the empirical relation between these fluxes is used. However, \citet{2009A&A...502..457G}  claim that the intrinsic X-ray luminosity, i.e., corrected by absorption, can also be used to estimate the MIR luminosity using:  $\log(L_{MIR}/1\times10^{43}$)\,=\,0.19\,+\,1.11 $\log(L_{2-10\,keV}$/1$\times10^{43}$). 
Both G05 and C14 have found similar values of the 2-10\,keV flux, 
$\sim$2.2$\,\times\,$10$^{−13}$\,erg\,
s$^{-1}$\,cm$^{-2}$. C14 reports an intrinsic luminosity 
of 8.71$\,\times\,$10$^{41}$\,erg\,s$^{-1}$, corrected from an absorption of 4$\,\times\,10^{23}$\,cm$^{-2}$, a value well below the CT limit. G05 only reports the contribution of the thermal component to the X-ray intrinsic luminosity, 4.1$\,\times\,$10$^{40}$\,erg\,s$^{-1}$, very low compared with value reported by C14 even if the value of the $N_H$ is higher in G05 fit, 1.6$\,\times\,$10$^{24}$\,cm$^{-2}$. 
This is due to the fact that the contribution of the thermal component in the 2-10\,keV is minimum as the peak temperature of the plasma is only 0.6\,keV. We have derived, using {\it Xspec} package, the intrinsic luminosity in the 2-10\,keV expected to arise from the AGN, i.e., associated to the power-law component of the G05 model. 
The resulting value, neglecting the contribution of the iron line and fixing the $N_H$ value to 1.6$\,\times\,$10$^{24}$\,cm$^{-2}$, is 8.6$\,\times\,$10$^{42}$\,erg\,s$^{-1}$, one order of magnitude above the luminosity calculated by C14. 
This is due to the fact that G05 and C14 analysis apply different values of $N_H$. For the analysis of C14, the value of $N_H$ (4$\,\times\,$10$^{23}$\,cm$^{-2}$) was derived fitting the spectrum, on the other hand, in the analysis of G05, the value of $N_H$ (1.6$\,\times\,$10$^{24}$\,cm$^{-2}$) was imposed assuming that Mrk\,622 is a CT object. 

For CT objects, $N_H$ is better determined when the X-ray spectrum reaches energies above 10\,keV, as the majority of the emission below this value is absorbed by the high Hydrogen column. A deep high X-ray observation (i.e., \textit{NuStar}) would be very useful to better estimate the value of $N_H$ for Mrk\,622. 

Using the relationship reported by \citet{2009A&A...502..457G} applied to the X-ray luminosity, results in an
estimated MIR luminosity of 1.4$\,\times\,$10$^{44}$\,erg\,s$^{-1}$, which can explain the observed \textit{CanariCam} emission, and at the same time allow us to reproduce the \textit{Spitzer} data using the same model shown in the upper panel of Figure~\ref{fig:Fig9}. Now, the contribution of the AGN in the MIR can explain the observed \textit{CanariCam} flux and we do not need to limit its contribution to the \textit{CanariCam} flux. 

Finally, we also consider the possibility that Mrk\,622 is variable in the X-rays. Many studies have shown that the X-ray emission of AGN can be highly variable. An order of magnitude variability in X-rays is frequently observed in AGN even in timescales of weeks \citep[i.e.,][]{2005ApJ...623L..93R}. If this is the case for Mrk\,622, it may be possible that the AGN was in a low state in X-ray during the epoch that the \textit{Spitzer} observation was performed (in 2005). In order to reproduce the observed MIR fluxed measured with \textit{CanariCam}, 79\,mJy, using the empirical relationship by \citet{2009A&A...502..457G}, an increment of the X-ray flux of one order of magnitude would be  necessary. Such variability changes in AGN are not uncommon. It is worth noting that the study of \citet{2015MNRAS.454..766A} was not carried out with simultaneous observations. However, this hypothesis can only be tested if X-ray and mid-IR simultaneous observations are available.

\section{Radio \textit{VLA} Data Analysis}
\label{VLA}

Faint Images of the Radio Sky at Twenty centimetres \textit{(FIRST)}  using the NRAO Karl G. Jansky Very Large Array \textit{(VLA)} report the detection of Mrk\,622 at 1.4 GHz \citep{1995ApJ...450..559B,1998AJ....115.1693C}
with an integrated flux of 9.0\,$\pm$\,0.5\,mJy. Data archives from the \textit{VLA},  Very-long-baseline interferometer \textit{(VLBI)}  and 
European VLBI Network \textit{(EVN)} were used to look for observations of this source.  So far, no \textit{VLBI} observations exist. 

\begin{figure}
\includegraphics[width=1.0\columnwidth,trim=0.0cm 3.5cm 0.0cm 1.5cm]{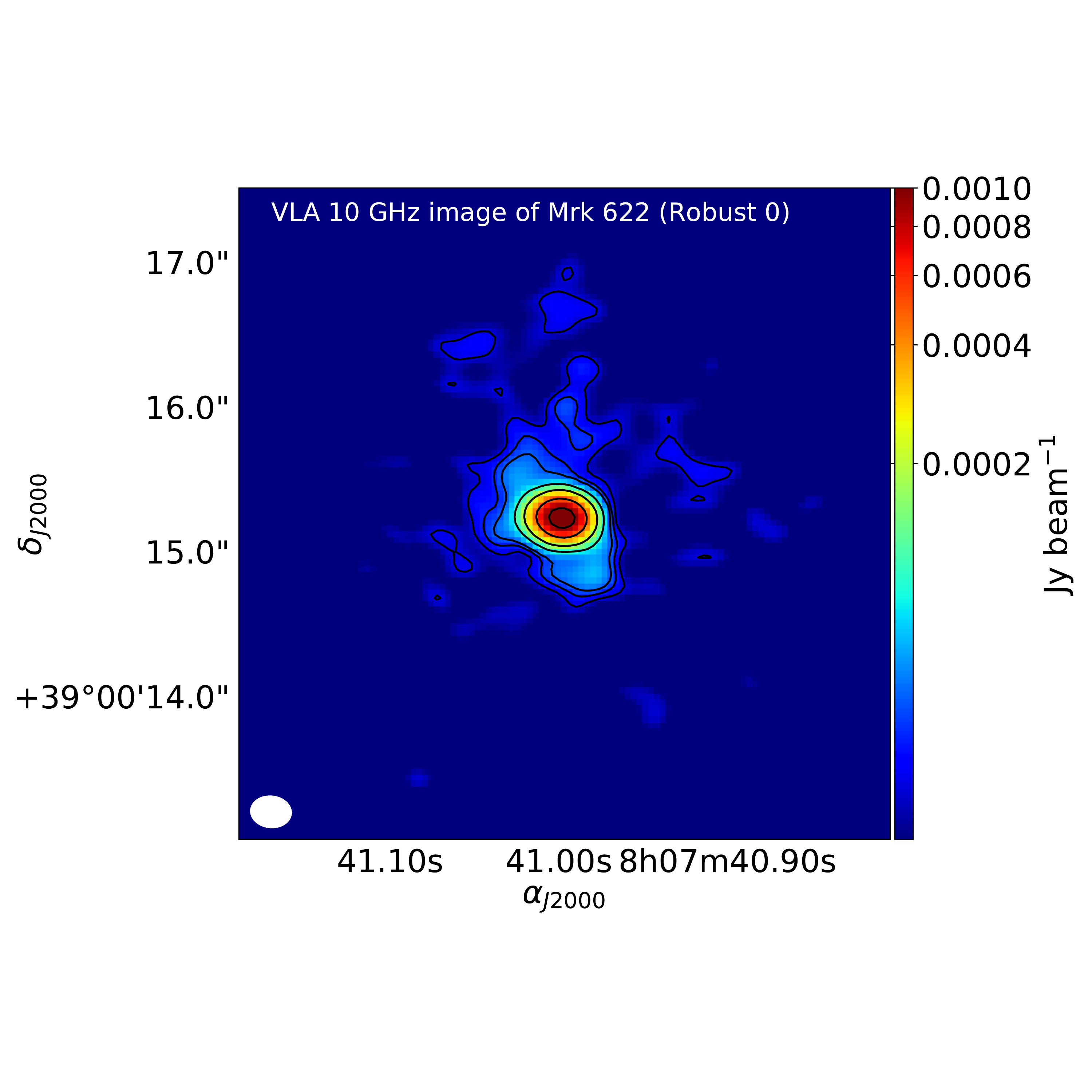}
\caption{\textit{VLA} map at 10\,GHz of Mrk\,622 obtained in March 6 2014 showing that the bright core source dominates the emission. The contours are at 0.03, 0.04, 0.050, 0.1, 0.2, 0.5, and 1.0\,mJy beam$^{-1}$. Note that 4$\sigma$ corresponds to 0.03\,mJy beam$^{-1}$.} 
\label{fig:vla1}
\end{figure}

\textit{VLA} observations at 10\,GHz obtained on 6 March 2014 in the A configuration with a spatial resolution of 0.2\arcsec were analysed. These observations were part of the project 13B-020. 
The resulting 10\,GHz \textit{VLA} image  is shown in Figure~\ref{fig:vla1}.

The morphology of the radio source can be described as a core with a total flux density of 1.47\,$\pm$\,0.03\,mJy, with two faint extensions displaced $\sim$ 0.5\arcsec to the NE and SW with total flux densities of 0.11\,$\pm$\,0.03 and 0.12\,$\pm$\,0.03 mJy, respectively, that extend beyond the central 1\arcsec core vicinity into a complex faint multiple source structure.

\begin{figure*}
\begin{center}
\subfigure[][]{%
\label{fig:vla2-a}%
\includegraphics[width=1.0\columnwidth,trim=0.0cm 3.5cm 0.0cm 1.5cm,height=8cm]
{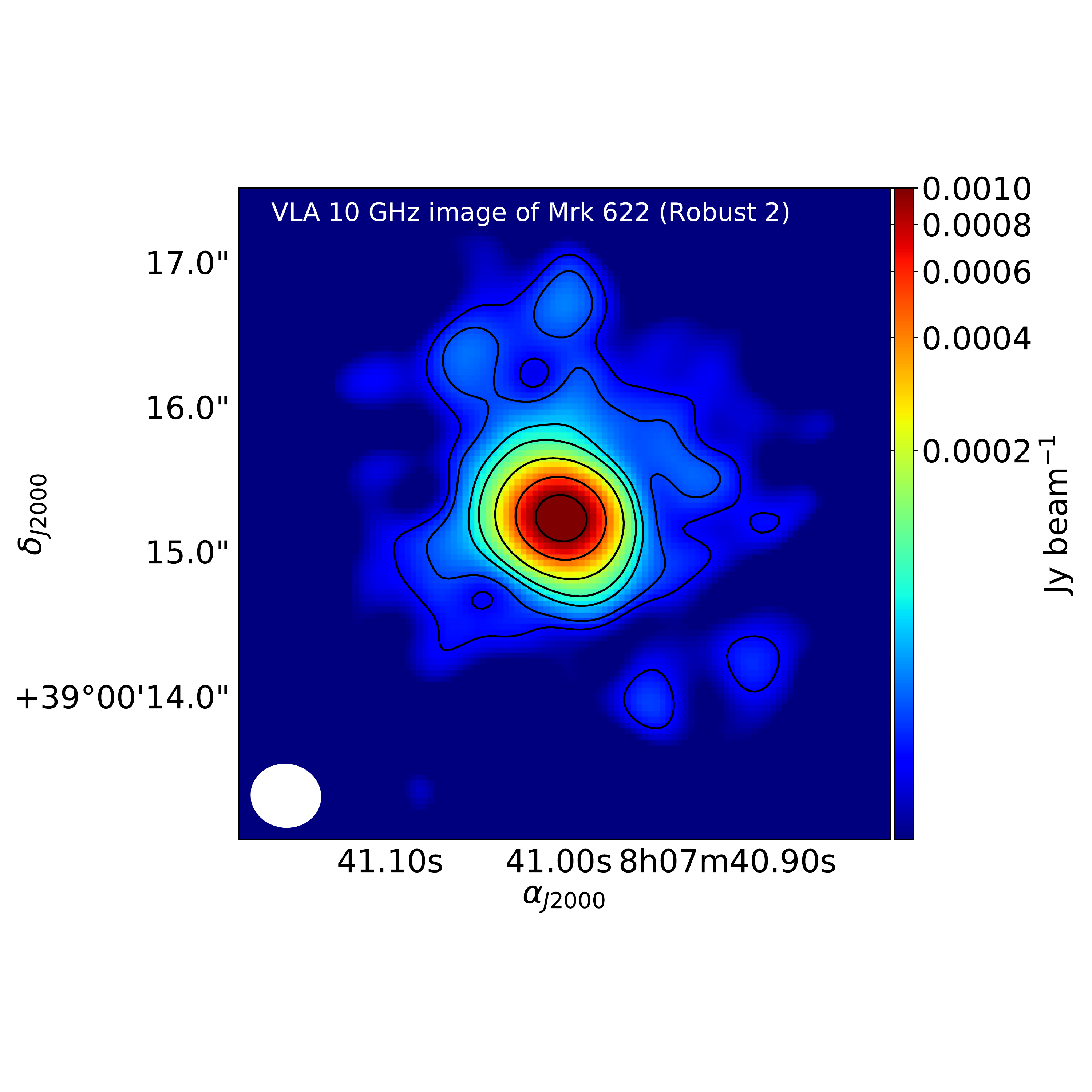}}%
\hspace{16pt}%
\subfigure[][]{%
\label{fig:vla2-b}%
\includegraphics[width=1.0\columnwidth,trim=0.0cm 2.5cm 0.0cm 1.5cm,height=7.7cm]
{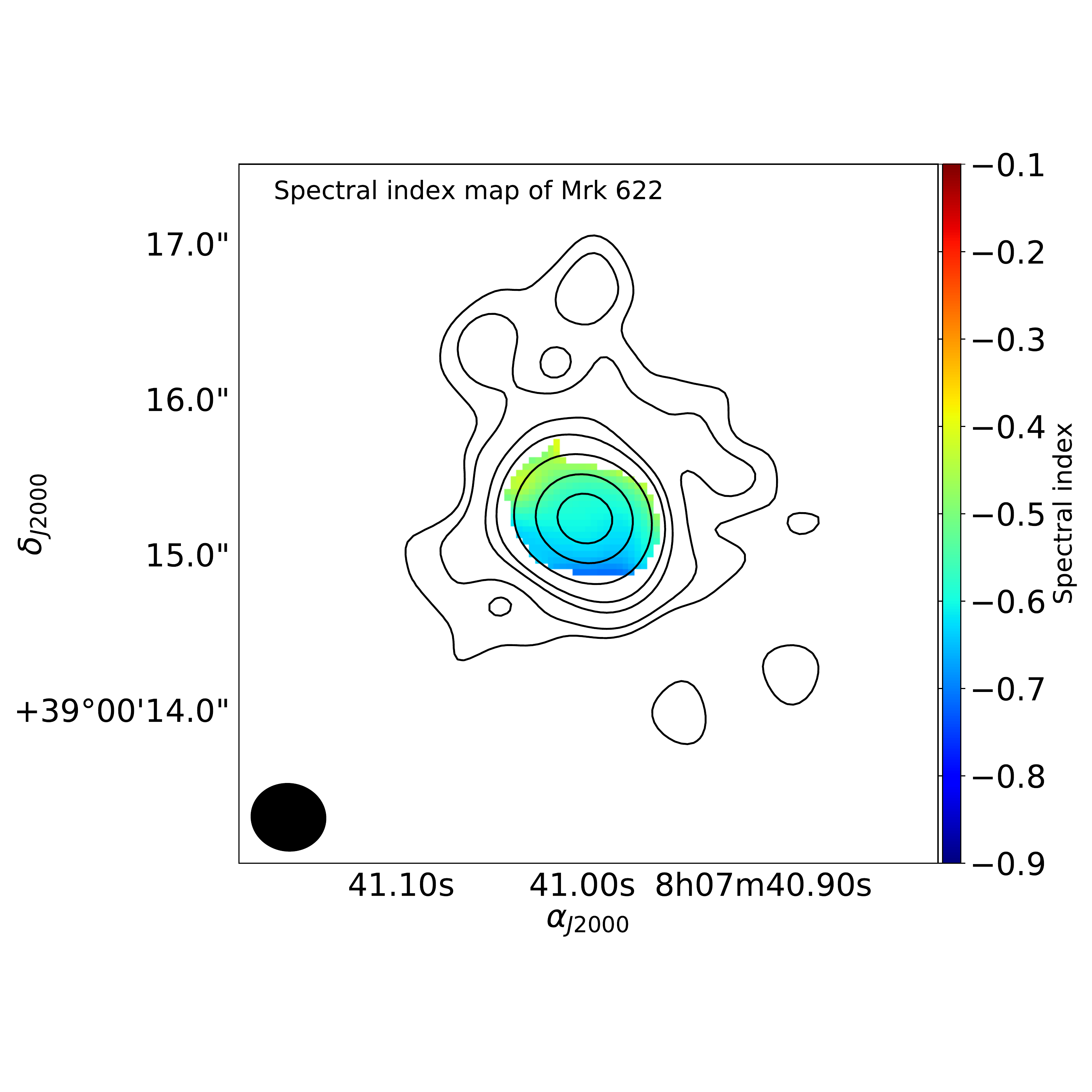}}%
\end{center}
\subfigure[][]{%
\label{fig:vla2-c}%
\includegraphics[width=1.0\columnwidth,trim=0.0cm 3.5cm 0.0cm 1.8cm,height=8cm]
{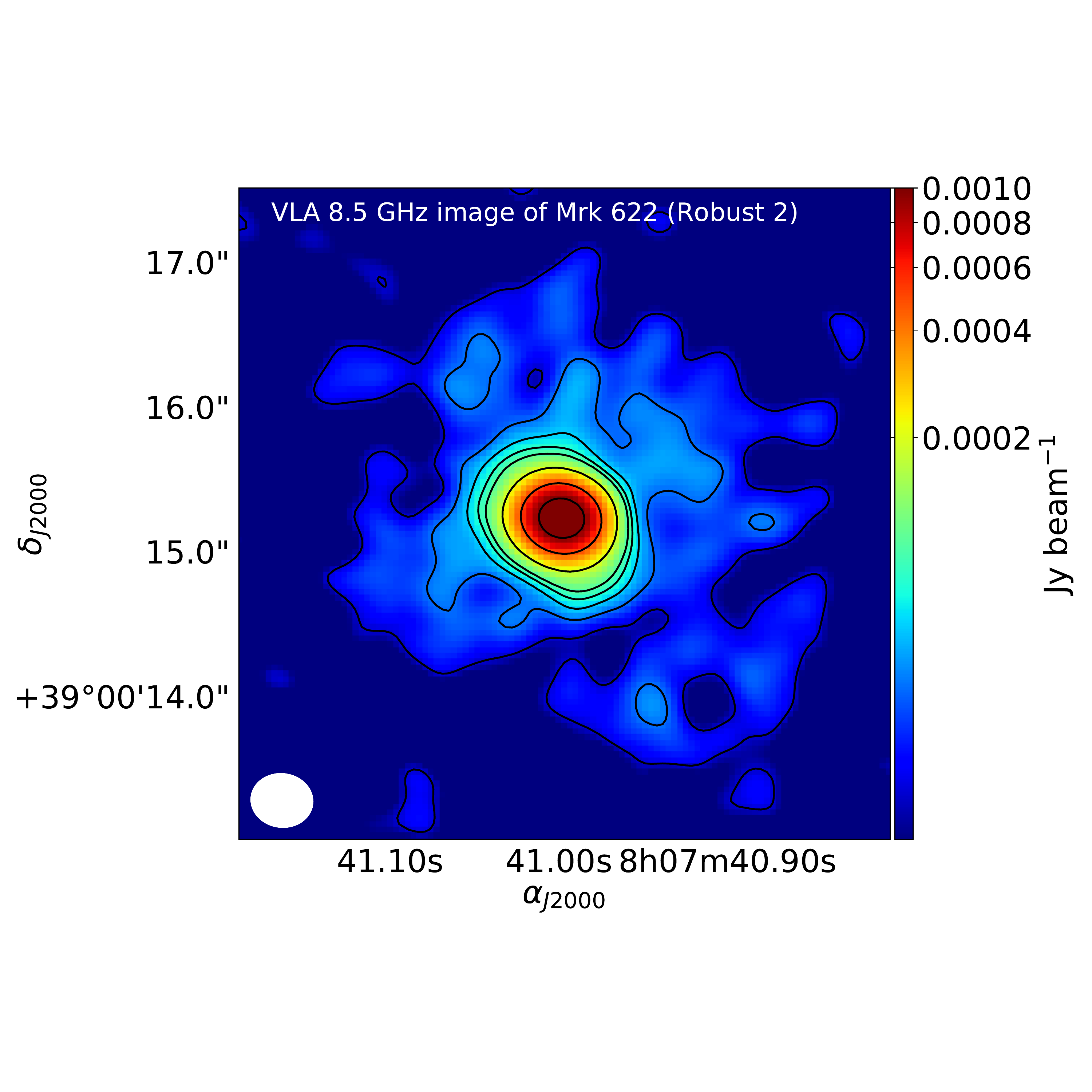}}%
\hspace{16pt}%
\subfigure[][]{%
\label{fig:vla2-d}%
\includegraphics[width=1.0\columnwidth,trim=0.0cm 3.5cm 0.0cm 1.8cm,height=8cm]
{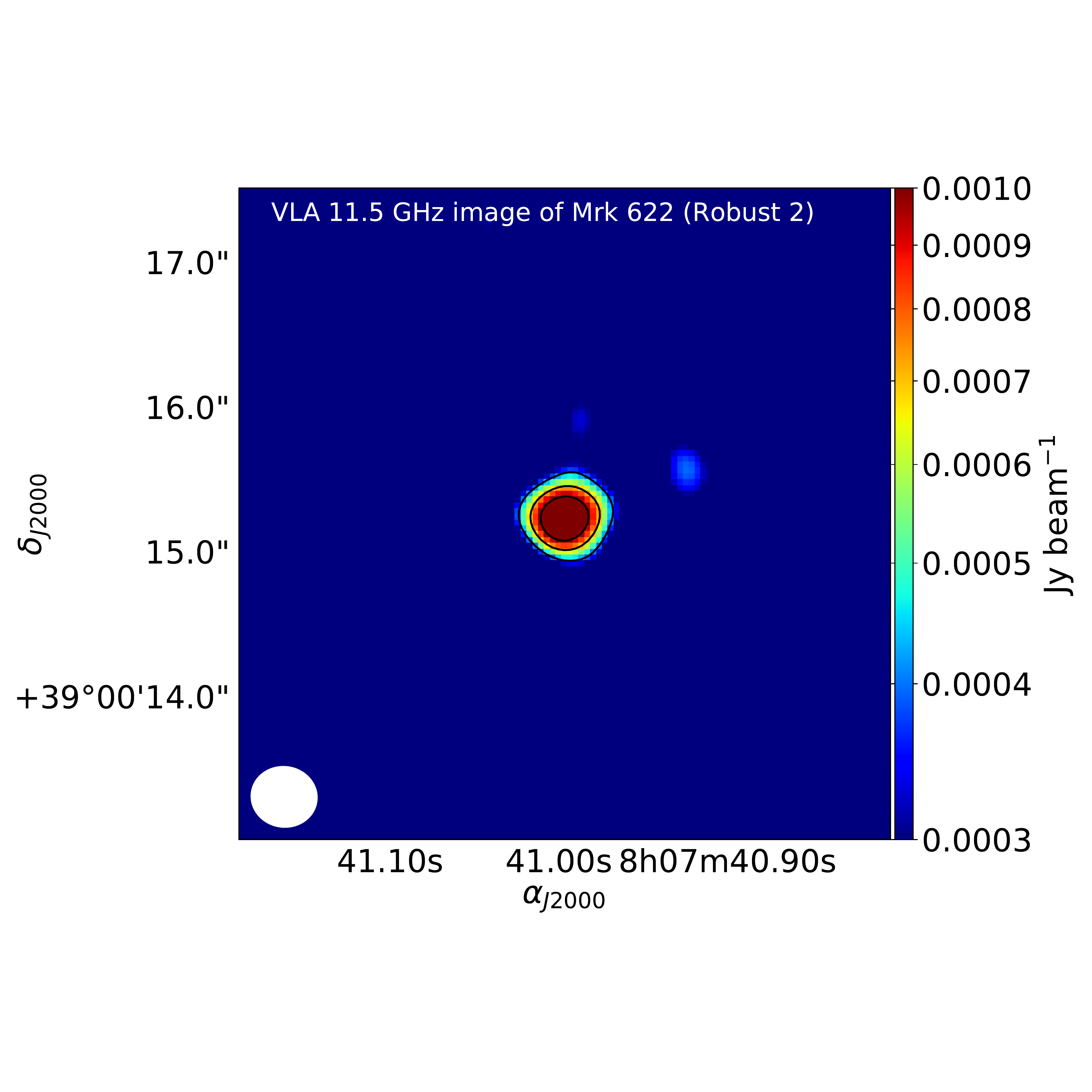}}
\caption{VLA images:
\subref{fig:vla2-a} made with the ROBUST parameter set to 2 (equivalent to natural weighting). This image has a synthesized beam of 0.28\arcsec\,$\,\times\,$\,0.22\arcsec; PA\,=\,84 deg. The contours are set at 0.03, 0.04, 
0.050, 0.1, 0.2, 0.5, and 1.0\,mJy beam$^{-1}$, respectively. The core is either extended along PA\,=\,70 degrees or contains multiple components along this PA;
\subref{fig:vla2-b} Spectral index map. The coloured bar represents the spectral index scale. The contours are the same as those in the intensity image map (c.f. \subref{fig:vla2-a});
\subref{fig:vla2-c} VLA image at 8.5\,GHz; and,
\subref{fig:vla2-d} VLA image at 11.5\,Ghz.
}
\label{fig:vla2}
\end{figure*}

The radio emission has been analyzed in order to confirm the possible binary nature of Mrk\,622 following the procedure proposed by \citet{2015ApJ...813..103M}.
The analysis is based on four criteria that helps to discriminate between dual-AGN, on one hand, and an AGN with an outflow or rotating disk on the other. These criteria are: radio morphology, 
size of the central bright sources, spectral index value of the compact sources, and spatial distribution of the spectral index (i.e., the spectral index image). 

Dual-AGN are confirmed when the morphology of the source is double (with each component tracing one of the SMBHs) and the components are unresolved (meaning that, within the errors, their size is comparable or smaller than the size of the beam). This second criterion helps to rule out the emission from extended regions located at the center of the studied object. Finally, flat or slightly steep spectrum radio sources ($\alpha \geq -0.8$) are interpreted as AGN, while  steeper spectra sources ($\alpha \leq -0.8$) emission in the surroundings can be interpreted as produced by jet components.

The VLA data of Mrk\,622 shows that the core is a central elongated structure that possible consists of more than one component. So it is not possible to apply the first criterium of \citet{2015ApJ...813..103M} to the VLA data. In order to resolve this region, data with more resolution are needed to confirm the existence of two or three components as was previously found in Paper\,I.

Spatially-smoothed images with CASA\footnote{https://casa.nrao.edu/ NRAO: Common Astronomy Software Applications} task ROBUST\,=\,2 (equivalent to natural weighting) and UVTAPER\,=\,300 klambda 
were produced. These parameters result in a synthesized beam of 0.28\arcsec\,$\,\times\,$\,0.22\arcsec and PA\,=\,84 deg. Using the CASA task CLEAN the Stokes I image as well as an image of the spectral index and an image of the spectral index error were obtained. The intensity image is shown in Figure~\ref{fig:vla2-a} and the spectral index map in Figure~\ref{fig:vla2-b}. 

Remarkably, the spatially smoothed intensity image shows structures that are not clearly detected in the 
image without smoothing (c.f., Figure~\ref{fig:vla1}). The core deconvolves to a source with dimensions 
of 82\,$\pm$\,13 mas $\,\times\,$ 41\,$\pm$\,20 mas, and a position angle of 70\,$\pm$\,18 degrees. This result suggests that the source is extended along a PA of $\sim$70 degrees or that it has multiple components along this PA. As can be seen in Figure~\ref{fig:vla2-a} faint double sources appear at about 1.5\arcsec~ to the NE and SW of the core. One possibility is that these external sources could be tracing the radio lobes of one or two jet systems, but a higher spatial resolution map is required to test this idea.  
In Figures~\ref{fig:vla2-c} and \ref{fig:vla2-d} the 8.5 and 11.5 GHz maps are presented. The map at 8.5\,Ghz shows the core source surrounded by a complex extended structure along PA = 70 deg, with the NE and 
SW multiple sources, while at 11.5 Ghz only the core is detected. The complex structure surrounding the core radio source  supports the previously proposed scenario that we are dealing with a double or multiple AGN system in Mrk\,622. 

The spatial distribution of the spectral index ($\alpha$ from $F_\nu \propto \nu^{\alpha}$) shown in Figure~\ref{fig:vla2-b}, include points with 
flux densities $>$\,30 $\mu$Jy and spectral index rms error better or $<$\,0.1. The inner parts of the image 
(a circle with a diameter of 1\arcsec) have a spectral index $\alpha$ of about -0.5\,$\pm$\,0.2, characteristic of AGN emission. The are surrounding this inner region shows a gradient in the spectral index from the NE to SW with values of $\alpha$ from -0.4 to -0.7. This is consistent with AGN jet studies \citep{2014AJ....147..143H,2015ApJ...813..103M}
which show flat core and steeper jet spectral indices, with flattening of $\sim$0.2 at the locations of the jet components due to particle acceleration or density enhancements along the jet. 

Unfortunately, the VLA images did not resolve the nuclear region, so in order to characterize the various faint components seen at 10 GHz higher spatial resolution observations are crucially needed.

\section{Results}
\label{res}

\begin{enumerate}

\item In order to verify the variability observed in the optical spectra of Mrk\,622, an estimation of the size of the NLR was done using the one-zone model of BL05. The NLR has a size of 2.7\,pc and the light-crossing time is 8.7\,yr. These estimations support that variability of the continuum flux and the NLR of Mrk\,622, if intrinsic to the source, can in occur in a time-span of 13 years. 
\smallskip
\item The variations detected in the doublet ratios $R\,=\,\frac{[SII]\lambda\,6716}{[SII]\lambda\,6731}$ of the central and red components suggest that the gas electron density is increasing in both the central and red components. 
\smallskip
\item In the \textit{WHT} spectrum, there is a hint of a displacement of the [OI]$\lambda$6300 red component with respect to the systemic velocity when comparison is made with the \textit{SDSS} spectrum  13 years earlier.

\smallskip

\item BL05 modeled the observed line ratio correlation found between [OIII]/H$\beta$ and [OIII]$\lambda$4363/$\lambda$5007. From the line ratios observed in Mrk\,622, the estimated electron density of the central component is  $\sim$3.5$\,\times\,$10$^{6}$\,cm$^{-3}$. Given such unusually high NLR density for a Type\,II AGN, temporal variations of the [OIII] lines could be explained if they originate from a region close to the nucleus.
\smallskip
\item The central, blue and red components occupy all the locus of strong AGN according to the WHAN diagram \citep[see][]{2011MNRAS.413.1687C} of line ratio $\log$[NII]$\lambda$6584/H$\alpha$ vs. $\log$ EW(H$\alpha$). This result is consistent with the BPT diagrams for the blue and red components. Although the central one is a composite SB+AGN using the BPT diagram, the WHAN diagram indicates that it is most probably a sAGN. 

\item Data from either the \textit{HST} and the \textit{GTC/CanariCam} images show that in spite of their relatively high spatial resolution, neither of them is capable of resolving the nuclear region of Mrk\,622. In passing, it is found that Mrk\,622 has a bar with a length of $\approx$\,3.64\,kpc, at position angle 74\,$\pm$\,1.5 degrees. The presence of a bar and a pseudobulge (S\'ersic index $n$\,=\,1.128$\pm$0.004) favors the scenario of secular evolution of the host galaxy.

\item Mrk\,622 has been observed by \textit{XMM}-Newton. From two different analysis found in the literature, although the measured value of the N$_H$ does not classify Mrk\,622 as a CT galaxy, it is remarkable that Mrk\,622 has a very high equivalent width of the Fe K$\alpha$ line ($\le$1600\,eV  or 900$^{+700}_{-600}$\,eV) depending on the analysis, an obscured AGN feature, indicating a possible CT nature of the source. Also the 2-10\,keV and [OIII] flux ratio indicates a CT nature, as reported by G05. Assuming that Mrk\,622 is a CT source, the 2-10\,keV X-ray luminosity associated to the AGN is 8.6\,$\,\times\,$\,10$^{42}$\,erg\,s$^{-1}$. 

\item  The \textit{IRS/Spitzer} spectrum of Mrk\,622 has also been retrieved. The relation between the 2-10\,keV band and the 12\,$\rm{\mu m}$ emission was used to test  whether the 12\,$\rm{\mu m}$ can be interpreted as originating in the same region as the X-ray emission. The result is positive only assuming that Mrk\,622 is a CT object, so the intrinsic X-ray luminosity is high enough to explain the whole \textit{CanariCam} flux of 79\,mJy  and fits well the \textit{IRS/Spitzer} spectrum. This is also true if Mrk\,622 is considered as a Compton-thin source and its X-ray flux would have suffered an increment of one order of magnitude at the moment of the IR observation with respect to the \textit{XMM-Newton} one.
\smallskip
\item Archival data at 10 GHz from the \textit{VLA} of Mrk\,622 were also analysed. The radio map shows a core source with two extensions to the SE and NW. When hyper resolving the image by adopting a uniform robust weighting of the \textit{VLA} data, the core deconvolves into a source with dimensions 82$\pm$13\,mas\,$\,\times\,$\,41$\pm$20\,mas, and position angle 70\,$\pm$\,18 degrees. This result suggests that the core is elongated or that it is constituted by multiple components distributed along an axis of $\sim$70 deg. At the distance of Mrk\,622 the core source is within a 34\,pc\,$\,\times\,$17\,pc region. The image also shows that after smoothing, two faint double sources appear at about 1.5\arcsec\, to the NE and SW of the core,  possibly associated with an AGN outflow. These faint sources were detected at 8.5 GHz, while at 13.5 GHz only the core source was detected. Furthermore, the spatial distribution of the spectral index $\alpha$ shows that within an aperture of 1\arcsec\, an AGN source is favoured since at the core $\alpha$\,=\,-0.5$\pm$0.2, and it steepens towards the extended sources, as would be expected for AGN jets.

\end{enumerate}

\section{Discussion}
\label{dis}

In Paper\,I, it is found that Mrk\,622 shows triple-peak emission lines. Also, it was shown that the spatial separation between the blueshifted and red-shifted components is $\sim$76\,pc. In this work, fits done to the triple-peaked profiles in both the \textit{WHT} and \textit{SDSS} spectra show that their positions in the WHAN diagram correspond to values, for the central, blue and red components, of $\log$\,[NII]/H$\alpha\,>$\,-0.4 and $EW$(H$\alpha$)$>$6~\AA, indicating that they all occupy the sAGN position. 

The extended emission regions (ENLR) measured in H$\alpha$ and [OIII]$\lambda$5007 have different sizes and morphologies. H$\alpha$ has an emission region with a size of 3.25\,kpc which may be the result of a SB-driven outflow. This kind of starburst is known to produce winds or bubbles of H$\alpha$ emission, as in NGC\,6240 \citep{2018Natur.556..345M}. The extended emission of [OIII]$\lambda$5007 has a size of 0.90\,kpc and could be the result of an AGN-driven outflow that produces a wind of ionised gas from the nuclear AGN. Since Mrk\,622 is a triple peaked source, it is also possible that a jet-driven outflow is responsible of the dynamics of the ionised gas in each of the three components. This scenario has been proposed to the nearby Sy\,2 galaxy NGC\,3393, a candidate previously claimed to be a binary AGN where the two SMBHs are separated by 150\,pc \citep[][]{2011Natur.477..431F}. This object is now proposed to harbor a pair of nuclear radio jets \citep{2018MNRAS.479.3892F}. 

In our analysis of the inner NLR, we assume photoionisation as the dominant ionisation mechanism, which we postulate might be caused by an earlier phase of intense AGN activity. Since \textit{no} AGN continuum nor a BLR are currently detected, we speculate that a \textit{delay} in light-travel propagation of the ionising continuum   might be the explanation. We propose that the current line variability might be caused by the propagation of an ionising front that has originated during an earlier phase of high AGN luminosity. An alternative possibility is that the emission results from shocks induced in the surrounding dense gas by relativistic jets. Either excitation mechanism requires a dense and very compact emission region otherwise, if the NLR gas was spherically distributed, the line intensity variations would be smoothed out over centuries rather than decades owing to light-travel time effects.

The optical spectra show that both the intensity of the NLR lines and the continuum have increased in brightness by about 45\% in a time-span of 13 years. The estimated radial extent of the NLR emission is 2.7\,pc (using the BL05 approach). Yet its geometrical distribution must be very compact and of unusually high density $\sim$3.5$\,\times\,$10$^{6}$ cm$^{-3}$ in order to account for the line variability. The estimated light-crossing time separating the active nucleus from the NLR gas is 8.7\,years. The origin of the variations require the propagation of an ionising front across the distinct NLR components.

The brightness surface analysis done for the HST NIR image of Mrk\,622 
shows that it 
has a pseudobulge, faint spiral arms and a bar-like structure. The analysis done in this work shows that the host galaxy can be classified as an (R)SB(rs)b in which a pseudo-ring has formed in the inner Lindblad resonance. \citet{1992A&A...257...17E} finds that barred-galaxies easily destroy rings or convert them in pseudo-rings due to interactions. However, results of the modelling of the NIR HST image suggest that the galaxy has already finished its relaxation process, since there is no evidence of tidal tails. 

On the other hand, since up to now there is only one observation in the X-ray bands of Mrk\,622, it is not possible to explore the possibility of having a CL AGN in the nucleus of Mrk\,622. The existing \textit{XMM}-Newton data show hints of CT nature of the source but they are not conclusive. There is an accepted observation project of Mrk\,622 with \textit{NuStar} but has not been observed yet. New X-ray observations of Mrk\,622 would allow us to investigate possible X-ray variation and also changes in line of sight obscuration, which could challenge its CT nature. 

Also, it is important to continue monitoring the variations observed in the optical region to see if these changes would imply a CL AGN. Due to its light crossing time of 8.7\,y, Mrk\,622 is a good candidate for future multi-wavelength variability monitoring programs.

The evolutionary scenario proposed for Mrk\,622 is that it has captured a dwarf galaxy and underwent a minor merger in the past, that was initially responsible for the triggering of an intermediate-luminosity AGN, i.e., with $L_{bol}=3.6\,\times\,10^{43}$\,erg\,s$^{-1}$. Such interactions are usually expected in spiral galaxies leaving intact the disk \citep{2001A&A...367..428A}. The morphology of the host galaxy shows structures typical of secular evolution, but there are also evidences that support the late minor merger scenario, in particular the presence of pseudo-rings and  the  probable CT nature of Mrk\,622. Recently, \citet{2017MNRAS.468.1273R} suggest that dual-AGN with projected distance separations of $<$10\,kpc are often found in Type\,II AGN due to huge amounts of obscuration carried towards the central region by a close galaxy encounter. Also, these authors demonstrate that the fraction of CT AGN is higher in late mergers.

The bolometric luminosity found analysing the X-ray (2-10\,keV) data is $L$\,=\,8.6$\,\times\,^{42}$\,erg\,s$^{-1}$,  i.e., $<$\,2\,$\,\times\,$\,10$^{44}$\,erg\,s$^{-1}$, therefore Mrk\,622 is an intermediate-luminosity CT AGN. These kind of AGN have a larger probability of being nuclear mergers \citep[see][]{2018Natur.563..214K}. So, if Mrk\,622 
underwent a minor merger that initially triggered the AGN and consequently left huge amounts of obscuring material, it is now in a late evolutionary phase. Since the MIR luminosity is $L_{MIR}$\,=\,1.4$\times10^{44}$erg\,s$^{-1}$, the obtained luminosity of Mrk\,622 is $\sim$10$^{10}$ L$_\odot$ which indicates that it is a luminous MIR source, indicative of an obscured AGN \citep{2014ARA&A..52..589H}. 

\begin{figure*}
\includegraphics[width=1.0\textwidth]{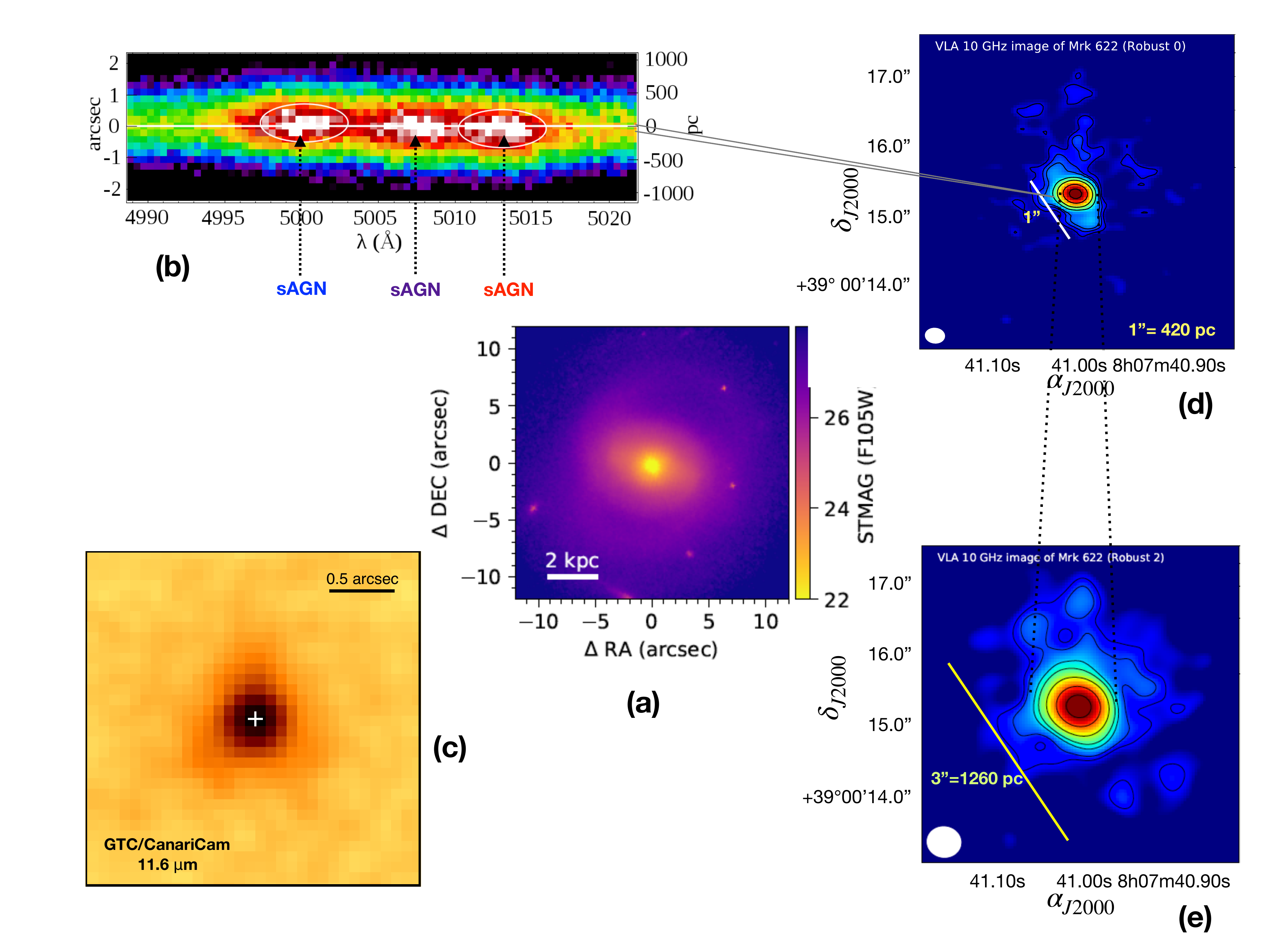}
\caption{Summary of the multi-wavelength observations of Mrk\,622. (a) At the center is the \textit{HST} WFC3 F105W image showing a bar of length $\sim$7.7\arcsec at  PA\,=\,74$^\circ$ and faint internal spiral arms.(b) 2-dimensional \textit{WHT}
long-slit spectrum (top-left) showing the triple peaked [OIII]$\lambda$5007 emission line: central, blue and red components, each one classified as strong AGN (sAGN) using the WHAN diagram. Note that the spatial separation of 0.16\arcsec\,$\sim$\,76 pc is between the red and blue components. (c) \textit{GTC/CanaryCam} image (bottom-left) at 11.6 $\mu$m of the unresolved ($<$0.4\arcsec$\,\times\,$0.34\arcsec) core source. (d) \textit{VLA} map (top-right) at 10\,GHz shows the central source and faint extensions located at 0.5\arcsec to the NE and SW from the core. (e) Intensity map (bottom-right) of the spatially-smoothed VLA, shows that the core deconvolves to a source of $\sim$82$\,\times\,$41 mas at PA\,=\,70$^\circ$ and multiple components along this PA, with faint sources about 1.5\arcsec to the SE and SW of the core. 
}\label{fig:multi}
\end{figure*}

In Figure~\ref{fig:multi} the spatial and structural results obtained in this multi-wavelength study are summarized. In  Figure~\ref{fig:multi}(a) the bar and faint spiral arms presented in Section~\ref{phot} are shown for comparison. In Figure~\ref{fig:multi}(b) the 2-dimensional \textit{WHT}
long-slit spectrum of the [OIII]$\lambda$5007 line, showing the central, blue and red peaks presented in Paper\,I is shown. Note the spatial scale in the Y-axis and the spatial separation of 0.16\arcsec or 76\,pc between the blue and red components. The three peaks are classified as sAGN in Section~\ref{class} using the WHAN diagram. The unresolved \textit{GTC/CanaryCam} 11.6 $\mu$m image is shown in Figure~\ref{fig:multi}(c) and discussed in Section~\ref{mir-obs}. On the right of Figure~\ref{fig:multi} are shown the \textit{VLA} maps at 10\,GHz (c.f., Sec.~\ref{VLA}) of the core source  and its faint SE and NW extensions at the top (Figure~\ref{fig:multi}(d)) and at the bottom (Figure~\ref{fig:multi}(e)) the spatially-smoothed core deconvolution into a
$\sim$82\,mas\,$\,\times\,$\,41\,mas and PA\,=\,70$^\circ$ (34\,pc\,$\,\times\,$\,17\,pc) core plus faint and  multiple components along the same  70$^\circ$  PA. It is worth to note that the separation of 76\,pc between the blue and red spectral components is in agreement with the size of the core \textit{VLA} source. 

Very long baseline interferometry (VLBI) observations are also required to resolve the morphology of the radio core source, but the possibility of  multiple components (the central one tracing an AGN) 
and four additional ones tracing one or two AGN inner jets would be of great interest. The {\it VLA} data at 10\,GHz show an elongated core AGN with a size of 82$\pm$13\,mas$\,\times\,$41$\pm$\,20\,mas and $PA\sim$\,70$^\circ$, which at the distance of Mrk\,622 corresponds to $<$\,4\,pc. So the AGN radio core is inside the central SMBHs influence radius r$_{G}$\,=\,4.3\,pc. 

The data analysed in this work does not allow us to completely rule out the uncoalesced binary SMBHs scenario for Mrk\,622, since the separation of the assumed binary SMBHs could be much smaller ($<$1\,pc) than what has been so far resolved, similar to what has been found for the Seyfert interacting galaxy NGC\,7674 (Mrk\,533) \citep[][]{2017NatAs...1..727K}. So, Mrk\,622 is a candidate for a sub-pc binary SMBHs system. 

\section*{Acknowledgements}
We thank the referee for making insightful suggestions and valuable comments that have greatly improved this manuscript. EB, ICG, JMRE, OGM, EJB, and CAN acknowledge support from DGAPA-UNAM grant IN113417. JMRE acknowledges support from the Spanish grants AYA2015-70498-C2-1, and AYA2017-84061-P de la AEI del Ministerio de Ciencia, Innovaci\'on y Universidades, Spain. OGM thanks support from DGAPA-UNAM grant IA103118. CAN thanks support from DGAPA-UNAM grant IN107313. EB, LB thank support granted by CONACYT. DRD acknowledges support from the Brazillian funding agency CAPES, via the PNPD program. LG thanks support from CONACYT project 167236. EJB acknowledges support from DGAPA-UNAM grant IN109217. LL acknowledges the financial support of DGAPA, UNAM grant IN112417, and CONACYT, Mexico. 
This work is based on observations made with the GTC, installed in the Spanish Observatorio del Roque de los Muchachos of the Instituto de Astrof\'isica de Canarias, in the island of La Palma. Also observations with \textit{WHT} operated on La Palma by the Isaac Newton Group in the Spanish Observatorio del Roque de los Muchachos of the Instituto de Astrof\'isica de Canarias. The National Radio Astronomy Observatory is a facility of the National Science Foundation operated under cooperative agreement by Associated Universities, Inc.

\bibliographystyle{mnras}
\bibliography{BenitezE} 

\bsp	
\label{lastpage}
\end{document}